\documentclass[useAMS,usenatbib]{mn2e}
\usepackage{amsmath}                                                            
\usepackage{amsfonts}
\usepackage{graphicx}
\usepackage{float}
\usepackage{morefloats}
\usepackage{listings}

\title[PRIMAL: Fitting to Gaia error added data]
{M2M modelling of the Galactic disc via PRIMAL: Fitting to Gaia error added data}
\author[J. A. S. Hunt \& D. Kawata]
 {Jason A. S. Hunt$^{1}$\thanks{E-mail: jason.hunt.11@ucl.ac.uk}
 and Daisuke Kawata$^{1}$
\\
$^{1}$ Mullard Space Science Laboratory, University College London,
Holmbury St. Mary, Dorking, Surrey, RH5 6NT, UK \\
}
\date{Submitted to MNRAS: $17^{th}$ December 2013.} 

\pagerange{\pageref{firstpage}--\pageref{lastpage}}
\pubyear{2012}

\begin{document}

\maketitle

\label{firstpage}

\begin{abstract}
We have adapted our made-to-measure (M2M) algorithm \sc{primal }\rm to use mock Milky Way like data constructed from an $N$-body barred galaxy with a boxy bulge in a known dark matter potential. We use M0 giant stars as tracers, with the expected error of the ESA space astrometry mission $Gaia$. We demonstrate the process of constructing mock $Gaia$ data from an $N$-body model, including the conversion of a galactocentric Cartesian coordinate $N$-body model into equatorial coordinates and how to add error to it for a single stellar type. We then describe the modifications made to \sc{primal }\rm to work with observational error. This paper demonstrates that \sc{primal }\rm can recover the radial profiles of the surface density, radial velocity dispersion, vertical velocity dispersion and mean rotational velocity of the target disc, along with the pattern speed of the bar, to a reasonable degree of accuracy despite the lack of accurate target data. We also construct mock data which take into account dust extinction and show that \sc{primal }\rm recovers the structure and kinematics of the disc reasonably well. In other words, the expected accuracy of the $Gaia$ data is good enough for \sc{primal }\rm to recover these global properties of the disc, at least in a simplified condition, as used in this paper.
\end{abstract}


\section{Introduction}
\label{intro-sec}
Making a computational model of the Milky Way is hardly a new concept, however neither is it a field which has reached its conclusion. There exist many mass models, which describe only the density distribution and the galactic potential \citep[e.g.][]{BS80,KZS02}. There exist kinematic models which describe the density and velocity distributions, but lack the constraint of self-consistent dynamics in the gravitational potential, such as the Besan\c{c}on model \citep[e.g.][]{Rea03}. Finally there exist dynamical models which satisfy this criteria \citep[e.g.][]{WPD08,B12-1}. Dynamical models can be constructed via different methods including Torus modelling \citep[e.g][]{McB12,MB13} and $N$-body modelling \citep[e.g.][]{GDRVZ13}. \cite{BR13} fit data from the Sloan Extension for Galactic Understanding and Exploration (SEGUE) and suggest the Milky Way's disc is maximal, in addition to constraining many dynamical properties of the disc. Models of the Milky Way are however always limited by the quality of the observational data they are based upon, and the more accurate data we have available to us, the better these models can become. A new generation of observational data about our Galaxy, unparalleled in both size and accuracy, is about to be produced by $Gaia$.

The European Space Agency's (ESA) $Gaia$ mission was launched on 19$^{th}$ December 2013 with an operational lifetime of 5 years, with provisions made for a possible 12-18 month extension. The estimated start of routine operations will be in 2014, with the first preliminary data release approximately 22 months after launch. The expected accuracy of $Gaia$ measurements is described for example in \cite{dB12}, with more detail on the astrometry in \cite{LLHOBH12}, the photometry in \cite{Jetal10} and the spectroscopy in \cite{Ketal11}. A large amount of preparatory software development and scenario modelling has already been occurring for the past decade \citep[e.g.][]{Ketal04,Wetal05,Setal11,LBSVBLS12,Rea12,APetal13,Brown2013,Xetal14}. The new wealth of information provided to us by $Gaia$ will need new methods to make the most of its potential. We are attempting to build a new dynamical model based on the made-to-measure (M2M) algorithm, ready for the $Gaia$ era.

Despite the significant increase in accuracy between $Gaia$ and previous surveys, e.g. Hipparcos, it will of course still be subject to error, due to both noise and calibration. The error will be dependent on stellar magnitudes, extinction and position in the sky. The astrometric parallax will carry the heaviest error and will in turn affect the error in the proper motions. The radial velocity error is heavily dependent on apparent luminosity and spectral type, but will be very accurate for red stars. 

The M2M method pioneered by \cite{ST96} has seen increasing interest in the last few years and has been used for multiple purposes. It has been applied to external galaxies \citep[e.g.][]{DeL07,DeL08,LM10,DGMT11,LM12,MG12,MGCMA13} and to the Milky Way itself \citep{BDG04,LMIII}. It has also been used to generate initial conditions for $N$-body models \citep{Deh09}. Despite its achievements so far, the M2M method still has many unexplored avenues of research open to us. 
\nocite{HKM13}
\nocite{HK12}

In Hunt \& Kawata (2013, hereafter Paper 1), and Hunt et. al. (2013, hereafter Paper 2), we have described the development of an M2M algorithm called \sc{primal }\rm (PaRtIcle-by-particle M2M ALgorithm). \sc{primal }\rm is designed to compare the observables at the position of each star, i.e. not binned data as in previous M2M modelling, because the Galactic stellar-survey data, such as the ones $Gaia$ will produce, are in the form of the position and velocity of individual stars. Another major difference between \sc{primal }\rm and other M2M algorithms is that the gravitational potential is calculated via self-gravity of the model particles. The potential is thus altered by the changing particle masses induced by the M2M algorithm. In Paper 1, we apply \sc{primal }\rm to the target system of a smooth axisymmetric disc created by $N$-body simulations and demonstrate that \sc{primal }\rm can reproduce the density and velocity profiles of the target system well, even when starting from a disc whose scale length is different from the target system. In Paper 2, we apply an updated methodology to disc galaxies with bar structure, and demonstrate that \sc{primal }\rm can reproduce the density and velocity profiles of these more complex targets, as well as providing a good estimate of the pattern speed of the bar.

In this paper, we first apply \sc{primal }\rm to the mock observational data of a single population of stars, M0III, which are constructed from a $N$-body simulated target galaxy. We ignore the dust extinction for simplicity and achieve a good recovery of the properties of the target system even with the $Gaia$ expected errors. Then we apply the dust extinction to the same mock target data and attain a reasonable recovery. Finally we apply extinction to mock data using Red Clump (RC) stars as tracers and compare the results for these different tracers.

This paper is organised as follows. Section \ref{Error} describes how we turn a target $N$-body galaxy model into mock observational data with $Gaia$ like errors. Section \ref{M2M} describes the M2M methodology of \sc{primal}\rm, with a more detailed explanation shown in Papers 1 and 2. Section \ref{R} shows the performance of our updated method for recreating the target disc system from the mock $Gaia$ data ignoring dust extinction to highlight the effects of the observational error. Section \ref{Ex} describes the results for the mock data taking the dust extinction into account. In Section \ref{SF} we provide a summary of this work.


\section{Target Setup}
\label{Target}

We use for demonstration a single target galaxy created with an $N$-body simulation. We selected our Target IV from Paper 2 as it shows boxy/peanut structure in the central bulge, which is thought to exist in the Milky Way \citep[e.g.][]{WG13}. It is set up using the method described in \cite{GKC12}, with the equations presented in Paper 2.

\begin{figure}
\resizebox{\hsize}{!}{\includegraphics{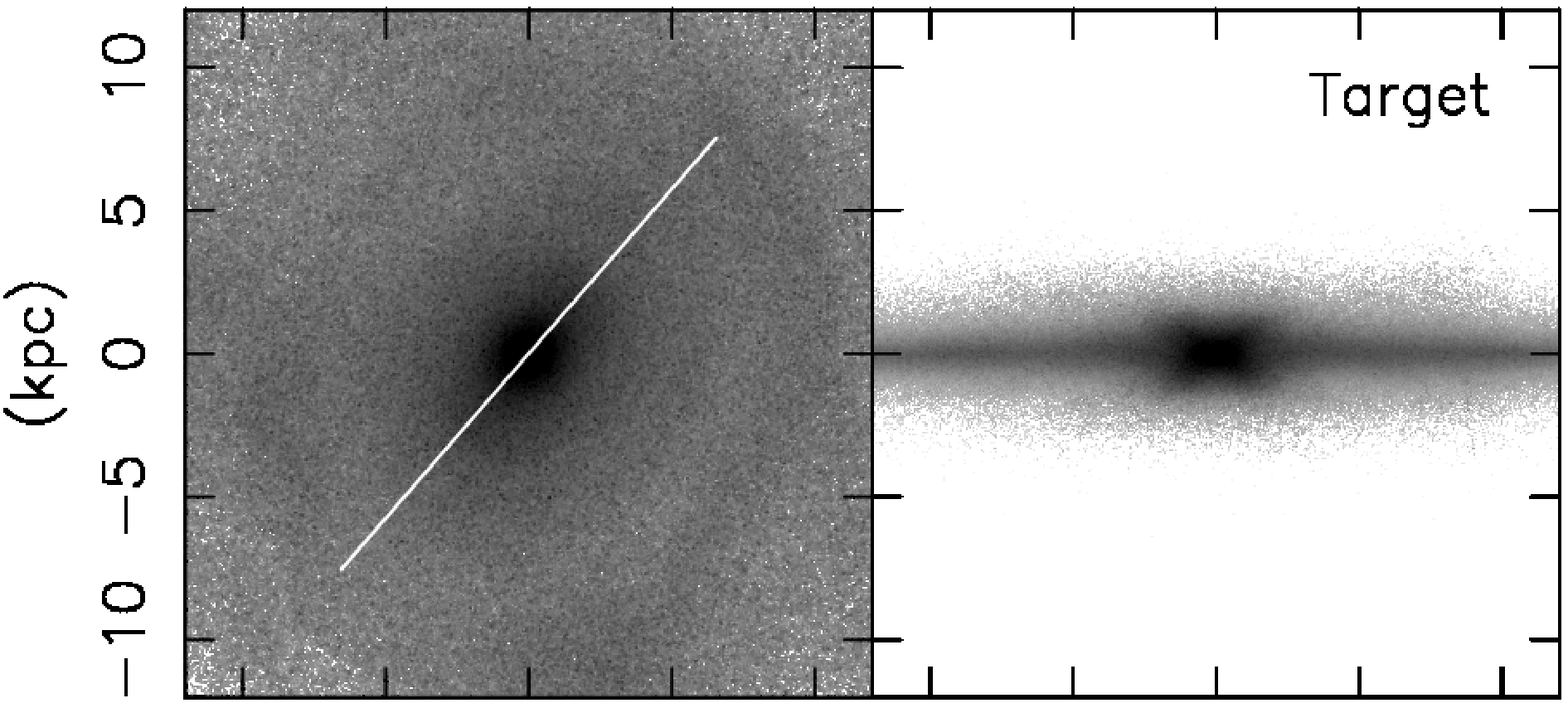}}
\resizebox{\hsize}{!}{\includegraphics{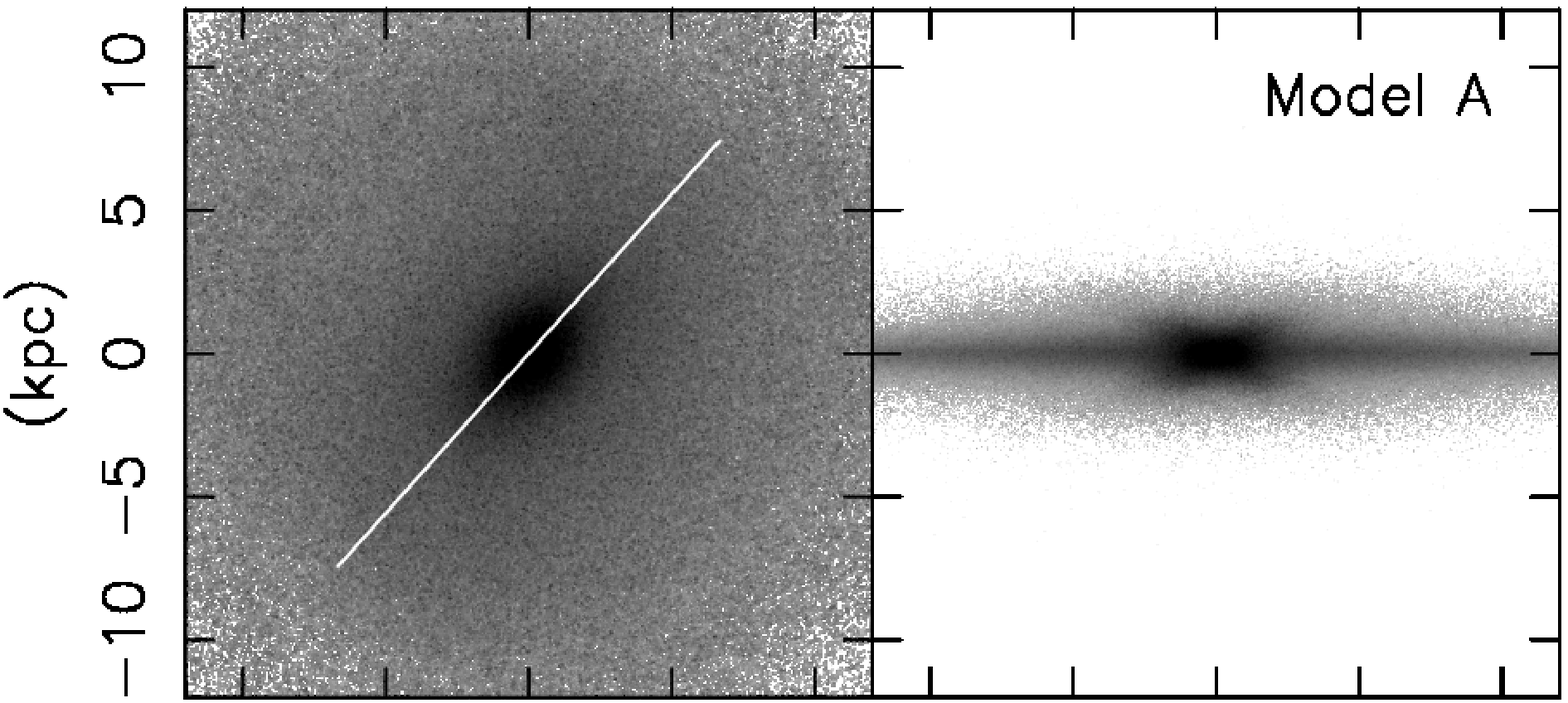}}
\caption{Face-on (left) and edge-on (right) density map of the Target (top) and Model A (bottom).}
\label{OT}
\end{figure}

The initial conditions for the target galaxy for this paper are constructed using the parameters $M_{200}=1.75\times10^{12}M_{\odot}$, $M_d=5.0\times10^{10}M_{\odot}$, $c=9.0$, $z_d=0.3$ kpc, $\sigma_r^2/\sigma_z^2=2.0$ and the scale length of the target disc is initially set as $R_{t,d}=3$ kpc as described in Paper 2. Our simulated target galaxy consists initially of a pure stellar disc with an exponential profile with no bulge and a static dark matter halo with the profile from \cite{NFW97}. We run an $N$-body simulation with these initial conditions, with $10^6$ particles, for 2 Gyr using a tree $N$-body code, GCD+ \citep{KG03,KOGBC13}, and adopt the final output as a target, shown in the top panel of Fig. \ref{OT}. We use the kernel softening suggested by \cite{PM07}. Although these authors suggested adaptive softening length, we use a fixed softening for these simulations for simplicity. Our softening length $\varepsilon=0.577\text{ kpc}$ is about three times larger than the equivalent Plummer softening length. We also use this softening for \sc{primal }\rm modelling runs.

For the model setup, as mentioned above, in this initial stage of development, we assume that the dark matter halo potential is known and there is no other external potential such as the bulge or stellar halo. We use the same number of particles, $10^6$, and the same dark matter halo and disc structure parameters for the model and target galaxies, except for the initial disc scale length: $R_d=2$ kpc for the models and $R_{t,d}=3$ kpc for the targets. We then evolve the model galaxy gravitationally while simultaneously adjusting it with \sc{primal}\rm.

\begin{figure}
\centering
\resizebox{\hsize}{!}{\includegraphics{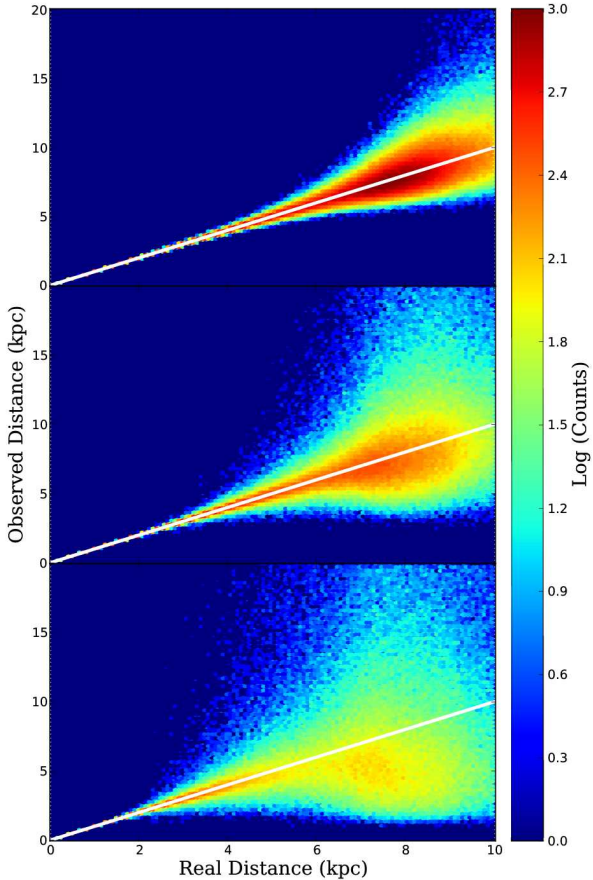}}
\caption{Real distance (1/$\pi$) compared to observed distance (1/$\pi_{obs}$) based on the $Gaia$ science performance estimates of the parallax for M0III stars without extinction (upper), M0III stars with extinction (middle) and RC stars with extinction (lower). The white lines lie along the 1:1 relation to guide the eye.}
\label{DE}
\end{figure}

\begin{figure*}
\centering
\resizebox{\hsize}{!}{\includegraphics{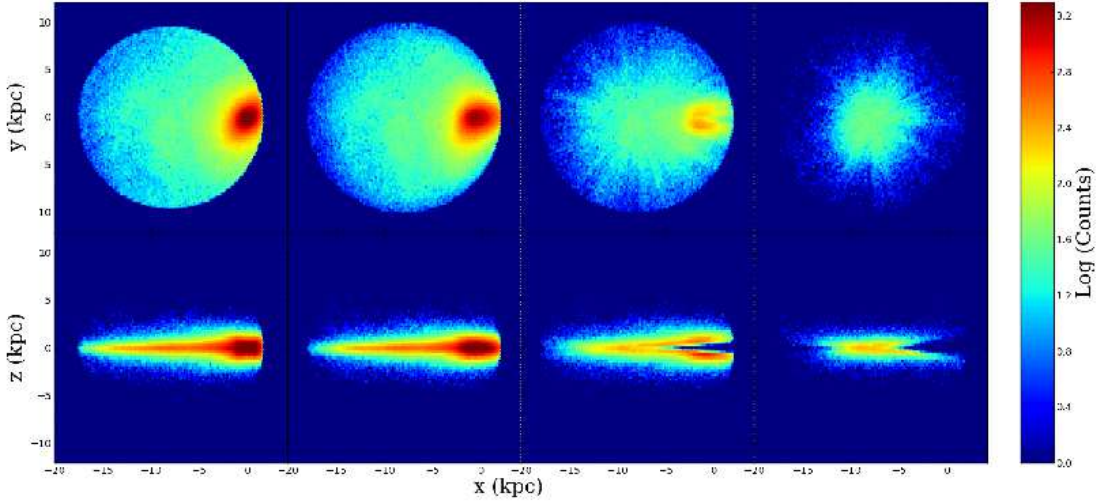}}
\caption{Face-on (upper) and edge-on (lower) logarithmic number counts of observed stars for M0III stars with no error (left) with $V\leq14.5$ mag, M0III stars with error but no extinction (middle left) with $V\leq14.5$ mag and $d_{obs}\leq10$ kpc, M0III stars with extinction (middle right) with $V\leq16.5$ mag and $d_{obs}\leq10$ kpc and RC stars with extinction (right) with $V\leq16.5$ mag and $d_{obs}\leq10$ kpc.}
\label{Exmap}
\end{figure*}


\section{Generating $Gaia$ mock data}
\label{Error}
Our target data are in Galactocentric Cartesian coordinates and hence must be converted into equatorial coordinates before we can add error based upon the $Gaia$ science performance estimates. 

First, we shift the centre to the solar position, with the orientation of the axes remaining unchanged. Then we change the orientation of the axes so that the x axis points in the direction of right ascension $\alpha=0^{\text{o}}$ and declination $\delta=0^{\text{o}}$, the y axis points in the direction of $(\alpha,\delta)=(90^{\text{o}},0^{\text{o}})$ and the z axis is aligned with the Earths North Pole using the transformation matrix, $\mathcal{T}$. We call this equatorial Cartesian coordinates. $\mathcal{T}$ is given by the inverse of the product of three rotation matrices, $\mathcal{T}=\mathcal{T}_1\mathcal{T}_2\mathcal{T}_3$, as shown in \cite{PCC03}.

$\mathcal{T}_1$ provides a rotation around the position angle of the North Celestial Pole with respect to the semi-circle passing through the North Galactic Pole and the zero Galactic longitude,
\begin{equation}
\mathcal{T}_1=\left[\begin{array}{ccc} \cos\theta_0 & \sin\theta_0 & 0 \\ \sin\theta_0 & -\cos\theta_0 & 0 \\ 0  &  0  & 1 \end{array} \right].
\end{equation}
$\mathcal{T}_2$ and $\mathcal{T}_3$ provide rotations around the equatorial position angles of the North Galactic Pole
\begin{equation}
\mathcal{T}_2=\left[\begin{array}{ccc} -\sin\delta_{\text{NGP}} & 0 & \cos\delta_{\text{NGP}} \\ 0 & -1 & 0 \\ \cos\delta_{\text{NGP}}  &  0  & \sin\delta_{\text{NGP}} \end{array} \right],
\end{equation}
and
\begin{equation}
\mathcal{T}_3=\left[\begin{array}{ccc} \cos\alpha_{\text{NGP}} & \sin\alpha_{\text{NGP}} & 0 \\ \sin\alpha_{\text{NGP}} & -\cos\alpha_{\text{NGP}} & 0 \\ 0  &  0  & 1 \end{array} \right].
\end{equation}
We use the values of $\theta_0=122.7^{\text{o}}$, $\delta_{\text{NGP}}=27^{\text{o}}27'$ and $\alpha_{\text{NGP}}=192^{\text{o}}49'30''$ for these angles, giving us:
\begin{equation}
\mathcal{T}=\left[\begin{array}{ccc} -0.0549 & -0.8734 & -0.4838 \\ 0.4941 & -0.4448 & 0.7470 \\ -0.8677  &  -0.1981 & 0.4560 \end{array} \right].
\end{equation}
The coordinate matrix, $\mathcal{A}$, for conversion from equatorial Cartesian coordinates to equatorial coordinates is given by:
\begin{equation}
\mathcal{A}=\left[\begin{array}{ccc} \cos(\alpha)\cos(\delta) & -\sin(\alpha) & -\cos(\alpha)\sin(\delta) \\ \sin(\alpha)\cos(\delta)  & \cos(\alpha) & -\sin(\alpha)\sin(\delta) \\  \sin(\delta) & 0 & \cos(\delta) \end{array} \right],
\end{equation}
such that;
\begin{equation}
\left[\begin{array}{c} v_r \\ \frac{k\mu_{\alpha}}{\pi} \\ \frac{k\mu_{\delta}}{\pi} \end{array}\right]    =\mathcal{A}^{-1}\mathcal{T}^{-1}\left[\begin{array}{c} U \\ V \\ W \end{array}\right],
\end{equation}
where $k=4.74$ is a unit conversion factor between the velocity of the star with respect to the Sun, $(U,V,W)$, in km s$^{-1}$ and the proper motions of the star $(\mu_{\alpha},\mu_{\delta})$ in arcsec yr$^{-1}$.

We treat the $N$-body particles as a single stellar population, which we will then add error to. We have chosen to use M0 giant (M0III) stars, with assumed $M_V=-0.4$ and $V-I_c=1.78$, for our tracers as these bright red giant stars will carry the least error in the estimation of their parallax and radial velocity. We assume each $N$-body particle (with $m_p=5\times10^4M_{\odot}$) corresponds to one M0III star, so there exists one M0 giant for every star cluster with mass $5\times10^4M_{\odot}$. This is a very simple assumption and does not follow a stellar population model or use a particular initial mass function (IMF). In reality, calculating the stellar mass density from the observed stars will be one of the biggest unknowns, because it is sensitive to their age, metallicity, IMF and evolutionary track. However, in this paper, we simply assume the conversion from M0III star number density to stellar mass density is known without any error, as a first step towards more realistic data. We will relax this strong assumption in future work.

We also assume we know the position and motion of the Sun. We locate the observer at $(-8,0,0)$ kpc in Fig. \ref{OT}, and the motion of the Sun is assumed to be $228.14$ km s$^{-1}$. Additionally in this paper, we generate error added data for any particle with $Gaia$ magnitude $G\leq20$ mag and $G_{RVS}\leq16.5$ mag. The relations to convert $V$ and $(V-I_c)$ to $G$ and $G_{RVS}$ \citep{Jetal10} are:
\begin{eqnarray}
G = &V& - 0.0257 - 0.0924(V-I_c) \nonumber \\ &-& 0.1623(V-I_c)^2  + 0.0090(V-I_c)^3,
\end{eqnarray}
and
\begin{eqnarray}
 G_{RVS} = &V& - 0.0119 - 1.2092(V-I_c) \nonumber \\ &+& 0.0188(V-I_c)^2 + 0.0005(V-I_c)^3.
\end{eqnarray}

We then add error to our target based upon the $Gaia$ performance estimates listed on the $Gaia$ website.\footnote{http://www.cosmos.esa.int/web/Gaia/science-performance}
A simple performance model, based upon the $Gaia$ Mission Critical Design Review, gives the equation for the end of mission parallax standard error, $\sigma_{\pi}$, as
\begin{equation}
\sigma_{\pi}=(9.3+658.1z+4.568-z^2)^{1/2}(0.986+(1-0.986)(V-I_c)),
\end{equation}
where
\begin{equation}
z=\text{max}(10^{0.4}(12-15),10^{0.4}(G-15)),
\label{zmax}
\end{equation}
and where $6\leq G\leq20$. 

For $6\leq G\leq12$, shorter integration times will be used to avoid saturating the CCDs. The end of mission performance will depend on the exact scheme used to avoid saturation, thus for the moment, equation (\ref{zmax}) allows us to ignore this uncertainty and returns a constant $\sigma_{\pi}=7$ $\mu$as for stars with $6\leq G\leq12$. We assume this same error for $G<6$, although $Gaia$ will not return data for $G<6$. Information on these very bright stars will be readily available from other surveys, and also the area covered by $G<6$ M0III stars will be covered by intrinsically fainter stars when using multiple populations. With M0III stars, $G=6$ corresponds to the apparent magnitude of stars at $d_{obs}\approx0.25$ kpc, and therefore only a small fraction of the mock data will be affected by this simplification.

The position and proper motion errors can be determined from a relationship with $\sigma_{\pi}$, which varies over the sky, and as such are derived from scanning law simulations. A table\footnote{http://www.cosmos.esa.int/web/Gaia/table-6} on the $Gaia$ Science Performance website shows the ecliptic longitude averaged numerical factor with which to multiply with $\sigma_{\pi}$, to return the appropriate value of $\sigma_{\alpha^{\ast}}$, $\sigma_{\delta}$, $\sigma_{\pi}$, $\sigma_{\mu_{\alpha^*}}$ or $\sigma_{\mu_{\delta}}$. This table$^2$ also takes into account the variation of the number of transits over the sky. 

Note that $\sigma_{\alpha^{\ast}}$ denotes the error in true arc, and may be converted to the standard right ascension with
\begin{equation}
\sigma_{\alpha^{\ast}}=\sigma_{\alpha} \cos(\delta),
\end{equation}
and similarly
\begin{equation}
\mu_{\alpha^{\ast}}=\mu_{\alpha} \cos(\delta).
\end{equation}
We then convert the proper motions to velocities in km s$^{-1}$ in the direction of $\alpha$ and $\delta$ with
\begin{equation}
v_{\alpha}=4.74(\mu_{\alpha}/\pi)\cos(\delta)
\end{equation}
and
\begin{equation}
v_{\delta}=4.74(\mu_{\delta}/\pi).
\end{equation}

However, because the error in the proper motions is also dependent on the error in the parallax, the errors must be convolved before they may be used in \sc{primal}\rm. We use the approximations
\begin{equation}
\sigma_{v_{\alpha}}=4.74\sqrt{\frac{1}{\pi^2}\left(\sigma_{\mu_{\alpha^*}}^2+\frac{\mu_{\alpha^*}^2}{\pi^2}\sigma_{\pi}^2\right)}.
\end{equation}
and
\begin{equation}
\sigma_{v_{\delta}}=4.74\sqrt{\frac{1}{\pi^2}\left(\sigma_{\mu_{\delta}}^2+\frac{\mu_{\delta}^2}{\pi^2}\sigma_{\pi}^2\right)}.
\end{equation}
to convolve the errors and also to convert the errors in $\mu_{\alpha^*}$ and $\mu_{\delta}$ to errors in $v_{\alpha}$ and $v_{\delta}$.

A simple performance model for the end of mission radial velocity error, $\sigma_{v_r}$, is given by:
\begin{equation} 
\sigma_{v_r} = 1 + b\text{e}^{a(V-14)},
\label{Vrerr}
\end{equation}
where $a$ and $b$ are constants dependent on the spectral type of the star. Some examples are given in a table\footnote{http://www.cosmos.esa.int/web/Gaia/table-5} on the $Gaia$ science performance website. This performance model is valid for $G_{RVS}\leq16.1$, where the fit error is 0.07 mag \citep{Jetal10}. The $a$ and $b$ values are estimated by linear interpolation as a function of $V-I_c$ using the table. We then apply these errors to the data from our M0III $N$-body target and displace the measured parallax, proper motion and radial velocity from the true values using random sampling.

Now that our data contain error, we need to strike a balance between the quantity of data available and the quality of the data, as stars with very large parallax errors provide incorrect information in the observables of our model. As such, we do not use all the available particles as points around which to calculate the observables, but merely those whose magnitude is within a predetermined limit. 

Fig. \ref{DE} shows the real distance from the observer compared to the observed distance for particles within 10 kpc for M0III stars (top). We first discuss a simplified case where the dust extinction effects are ignored. The effects of the dust extinction will be discussed in Section \ref{Ex}. The observed distance, $d_{obs}$, in Fig. \ref{DE} is the error added distance from the observer. $d_{obs}$ is calculated from the randomly displaced parallax measurement, $\pi_{obs}$, following the expected parallax errors. The top panel of Fig. \ref{DE} shows that the accuracy of the distance measurement is excellent within 4 kpc, but starts to diverge quickly at higher distances. It also shows that while the difference between the observed and correct positions for the majority of stars remain within $\approx2$ kpc even up to $d=10$ kpc, a significant fraction have errors of more than $50\%$. For this paper we have set the limit for the selection of the data to be $d_{obs}<10$ kpc. We also add the selection limit of $V\leq14.5$ mag for obtaining accurate radial velocities. Note that this estimate of distance error uses only parallax distance estimates, whereas from the real $Gaia$ data it is also possible to measure photometric distances which may help to reduce the error.

Fig. \ref{Exmap} shows the face-on (upper panels) and edge-on (lower panels) distribution of generated M0III stars which meet our selection criteria ($V\leq14.5$ mag and $d_{obs}\leq10$ kpc). The left panels show the true distribution of the selected stars and the second column shows the distribution of the stars after the error has been added, i.e. the position of the stars after the random displacements in parallax. Fig. \ref{Exmap} shows the target data reaches the centre of the galaxy. However the observed shape of the bar differs between the true distribution and the error added data. With the addition of error, the boxy structure of the bar is much weaker and the angle of the bar becomes less apparent.

\begin{table*}
\centering
\caption{M2M model results at the final timestep. $\Omega_p$ is the model pattern speed, with a target of 28.9 $\text{km s}^{-1}\text{kpc}^{-1}$, $\chi^2_{\rho}$ is a measure of accuracy of the density, $-\mathcal{L}_{r,v_{\alpha},v_{\delta}}$ are the likelihood values for the radial velocity and proper motions.}
\label{Mpars}
\renewcommand{\footnoterule}{}
\begin{tabular}{@{}cccccccccccc@{}}
\hline
Model & $\Omega_p$ ($\text{km s}^{-1}\text{kpc}^{-1}$) & $\chi^2_{\rho}$ & $-\mathcal{L}_{v_r}/10^6$ & $-\mathcal{L}_{v_{\alpha}}/10^6$ & $-\mathcal{L}_{v_{\delta}}/10^6$ & Number of tracers selected & Notes \\ \hline
$i$ & 34.3 & 0.370 & 7.831 & 8.2112 & 8.2012 & 558,852 & Unconstrained \\ \hline
A & 28.5 & 0.100 & 5.832 & 5.8898 & 5.8288 & 558,852 & No Error \\ \hline
B & 28.6 & 0.137 & 7.067 & 2.6357 & 2.6315 & 517,527 & Fiducial Model \\ \hline
C & 26.1 & 0.126 & 6.836 & 2.6365 & 2.6322 & 517,527 & $\rho$ only \\ \hline
D & 25.0 & 0.130 & 6.926 & 2.6363 & 2.6322 & 517,527 & $\rho$ and $v_r$ only \\ \hline
E & 33.8 & 0.130 & 6.939 & 2.6356 & 2.6314 & 517,527 & $\rho$ and $v_{\alpha,\delta}$ only \\ \hline
F & 22.5 & 0.196 & 6.885 & 2.6364 & 2.6332 & 517,527 & $\rho\in$ $V\leq14.5$ mag \\ \hline
G & 25.9 & 0.196 & 8.815 & 2.6362 & 2.6327 & 517,527 & $R_{d,ini}=4$ kpc \\ \hline
G$_i$ & 23.9 & 0.274 & 8.898 & 2.6373 & 2.6360 & 517,527 & Model G unconstrained \\ \hline
H & 25.6 & 0.167 & 7.548 & 2.6375 & 2.6332 & 517,527 & $M_{d,ini}=10^{11}M_{\odot}$ \\ \hline
H$_i$ & 44.4 & 3.624 & 10.99 & 4.0241 & 4.0203 & 517,527 & Model H unconstrained \\ \hline
I & 27.7 & 0.249 & 1.666 & 0.8222 & 0.8211 & 173,821 & M0III with extinction \\ \hline
J & 27.3 & 1.593 & 0.442 & 0.2399 & 0.2396 & 52,111 & RC with extinction \\ \hline
\end{tabular}
\end{table*}


\section{The M2M Algorithm: PRIMAL}
\label{M2M}

The M2M algorithm works by calculating observable properties from the model and the target, and then adapting particle masses such that the properties of the model reproduce those of the target. The target can be in the form of a distribution function, an existing simulation or real observational data. The model can be a test particle simulation in an assumed fixed or adaptive potential, or a self-gravity $N$-body model.
 
We have presented a full description of both the original M2M and our particle-by-particle M2M in Papers 1 and 2. In this section, we describe briefly the basis of our particle-by-particle M2M. As mentioned in Section \ref{intro-sec}, our ultimate target is the Milky Way, where the observables are not binned data, but the position and velocity of the individual stars which are distributed rather randomly. To maximise the available constraints, we evaluate the observables at the position of each star and compare them with the $N$-body model, i.e. in a particle-by-particle fashion. To this end, \sc{primal }\rm uses a kernel often used in Smoothed Particle Hydrodynamics (SPH), $W(r,h)$, which is a spherically symmetric spline function given by
\begin{equation}
\begin{array}{l}
W(r,h) = \frac{8}{\pi h^{3}} 
 \times \left\{ \begin{array}{cc}
 1-6(r/h)^{2}+6(r/h)^{3} & {\rm if}\ 0\leq r/h\leq 1/2,  \\
 2[1-(r/h)]^{3}      & {\rm if}\ 1/2\leq r/h\leq 1,  \\
 0               & {\rm otherwise},
\end{array} \right.\\
\end{array} 
\label{Weq}
\end{equation}
as shown in \cite{ML85}, where $r$ is the distance to the neighbour particle and $h$ is a smoothing length described later. Note that in \sc{primal}\rm, the kernel, $W(r,h)$, does not explicitly include the total mass, $M_{\text{tot}}$, unlike standard M2M algorithms, because we wish to eventually apply it to the Milky Way, whose mass is unknown. 

In a change from Paper 1 and 2, we have converted the algorithm to take target data in equatorial coordinates, e.g. right ascension, $\alpha$, declination, $\delta$, parallax, $\pi$, radial velocity from the position of the Sun, $v_r$, and proper motions, $v_{\alpha}$ and $v_{\delta}$. We make this change as this is the form in which $Gaia$ will return its data. We maintain six dimensional phase space information, and as such no accuracy should be lost at this stage.

We again convert our galactocentric Cartesian model data into equatorial coordinates to compare the radial velocity and proper motion observables constructed from the $Gaia$ data via the process shown in Section \ref{Error}. We then calculate the velocity likelihood observables in equatorial coordinates, using the equations derived in Paper 2, e.g. for $v_{\alpha}$, the likelihood is given by:
\begin{equation}
\hat{\mathcal{L}}_{v_{\alpha,j}}=\frac{1}{\sqrt{2\pi}}\sum_i W_{ij}m_i\text{e}^{-(v_{\alpha,j}-v_{\alpha,i})^2/2\sigma_{v_{\alpha},j}^2},
\end{equation}
for model particle $i$ and target particle $j$.

We also convert the target particle positions into Cartesian coordinates to allow the same form of density observable as Papers 1 and 2, using the equation:
\begin{equation}
\left[\begin{array}{c} x \\ y \\ z \end{array}\right]=\mathcal{T}\left[\begin{array}{c} \cos(\alpha)\cos(\delta)/\pi \\ \sin(\alpha)\cos(\delta)/\pi \\ \sin(\delta)/\pi\end{array}\right],
\end{equation}
using the observed parallax, $\pi_{obs}$, as discussed in Section \ref{Error}. We then use the same density observable as Papers 1 and 2 for both the target and the model. For example for the target;
\begin{equation}
\rho_{t,j}=\sum^N_{k=1}m_{t,k}W(r_{kj},h_j),
\end{equation}
where $m_{t,k}$ is the mass of the target particle, $r_{kj}=\vert\textbf{r}_k-\textbf{r}_j\vert$, and $h_j$ is the smoothing length determined by 
\begin{equation}
h_j=\eta\left(\frac{m_{t,j}}{\rho_{t,j}}\right)^{1/3},
\label{smoothing}
\end{equation}
where $\eta$ is a parameter controlling smoothness which we have set to $\eta=3$.  In SPH simulations, a value of $\eta$ between 2 and 3 are often used, and we employ the relatively higher value to maximise the smoothness. This results in $\approx113$ neighbouring particles being included in the smoothing when the particles are distributed homogeneously in three-dimensional space. The solution of equation (\ref{smoothing}) is calculated iteratively until the relative change between two iterations is smaller than $10^{-3}$ \citep{PM07}.

Note that the positions of the target stars are displaced due to the parallax errors, and the observables $\rho_{t,j}$ do not correctly represent the density of the target system. Our target stars are selected with $V\leq14.5$ mag, and the observed distance $d_{obs}\leq10$ kpc, as mentioned in Section \ref{Error}. We do however include particles with $V>14.5$ mag and $d_{obs}>10$ kpc in the calculation of the density observables themselves. This helps to compensate for the underestimation of the density of the target stars just inside of the magnitude cut, for which there are significant number of stars fainter than the magnitude cut, but within the smoothing length. However this also counts fainter stars whose observed distance is much smaller than the real distance due to the error which can result in overestimation of the local density.

\begin{figure*}
\centering
\resizebox{\hsize}{!}{\includegraphics{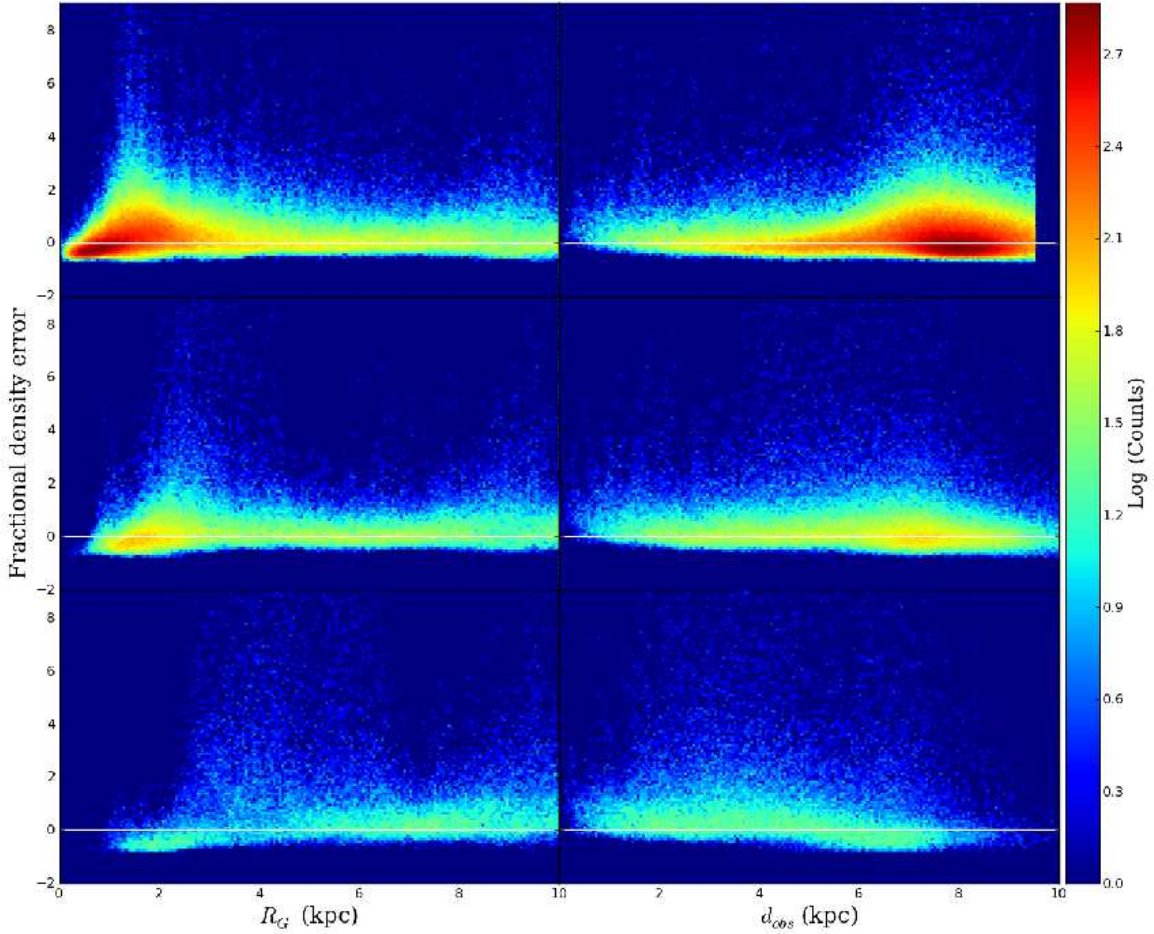}}
\caption{Fractional density error $(\rho_{\text{obs}}-\rho_{\text{true}})/\rho_{\text{true}}$ as a function of observed Galactocentric radius, $R_G$ (left) and observed distance from the Sun, $d_{obs}$ (right), for M0III stars without extinction (upper), M0III stars with dust extinction (middle) and RC stars with dust extinction (lower), coloured by logarithmic number density of the stars. The white line lies along zero to guide the eye.}
\label{DenError}
\end{figure*}

Fig. \ref{DenError} shows the fractional density error of the mock data against Galactocentric radius for M0III stars (upper left). The upper left panel of Fig. \ref{DenError} shows density tends to be overestimated when using this simplistic calculation of the density. Most notably the panel shows a substantial overestimation between 1 and 2 kpc from the galactic centre. This overestimation can be understood from the face-on view of the distribution of stars shown in the upper panels of Fig. \ref{Exmap}. In the data with the true particle positions (left) the bar is clearly shown. On the other hand, in the error added data (2nd column) the bar shown is more diffuse, and the apparent angle of the bar looks different. Therefore, for example, while (x,y)=$(-2,0)$ is the edge of the bar in the true distribution, because of the large errors in parallax, the observed distance of many stars in the bar are randomly displaced from the true bar location, which makes the bar appear more diffuse. As a result, the density at (x,y)=$(-2,0)$ increases, which leads to the overestimation seen at $R_G\approx2$ kpc in Fig. \ref{DenError}. Fig. \ref{DenError} also shows an underestimation in the inner 0.5 kpc region. This is also understandable from Fig. \ref{Exmap}, for the same reason, because the observed central concentration is more diffuse due to the large parallax error at the centre, the very centre of the galaxy appears less dense. In this paper we simply take the measured density. However, because of our particle-by-particle M2M algorithm, we have many target stars, and demonstrate that \sc{primal }\rm works reasonably well even with this simple density measurement.

Fig. \ref{DenError} shows the fractional density error of the mock data against observed distance for M0III stars without extinction (upper right). There is a general trend of overestimation matching that which is seen in the upper left panel. The cut off of the data at $d_{obs}\approx9.55$ kpc is due to the magnitude limit of $V\leq14.5$ mag for the data selection.

Similarly, the density of the model at the target position is calculated with
\begin{equation}
\rho_{j}=\sum_{i=1}^{N}m_{i}W(r_{ij},h_{j}).
\label{rho}
\end{equation}
The target density $\rho_{t,j}$ is calculated only once at the beginning of the M2M simulation, and the model density $\rho_{j}$ is recalculated at every timestep. We then calculate the difference between the density observables thus
\begin{equation}
\Delta_{\rho_j}(t)=\frac{\rho_j(t)-\rho_{t,j}}{\rho_{t,j}}.
\end{equation}
Having converted the observables into their appropriate coordinates, we then compare these observables with the same method as Paper 2, resulting in the change of mass equation: 
\begin{eqnarray}
&\frac{d}{dt}m_{i}(t)& = -\epsilon m_{i}(t)\Biggl\{M\sum_{j} \frac{W(r_{ij},h_j)}{\rho_{t,j}}\Delta_{\rho_j}(t) \nonumber \\  
&-& \zeta M\Biggl[\sum_{j}W_{ij}\left(\frac{1}{\sqrt{2\pi}}\frac{\text{e}^{-(v_{r,j}-v_{r,i})^2/2\sigma^2_{v_r,j}}}{\hat{\mathcal{L}}_{v_r,j}}-\frac{1}{\rho_j(t)}\right) \nonumber \\
&+& \sum_{j}W_{ij}\left(\frac{1}{\sqrt{2\pi}}\frac{\text{e}^{-(v_{\alpha,j}-v_{\alpha,i})^2/2\sigma^2_{v_{\alpha},j}}}{\hat{\mathcal{L}}_{v_{\alpha},j}}-\frac{1}{\rho_j(t)}\right) \nonumber \\
&+& \sum_{j}W_{ij}\left(\frac{1}{\sqrt{2\pi}}\frac{\text{e}^{-(v_{\delta,j}-v_{\delta,i})^2/2\sigma^2_{v_{\delta},j}}}{\hat{\mathcal{L}}_{v_{\delta},j}}-\frac{1}{\rho_j(t)}\right)\Biggr] \nonumber \\
&+&  \mu \left(\ln \left(\frac{m_{i}(t)}{\hat{m}_{i}}\right)+1\right)\Biggr\},
\end{eqnarray}
where $\hat{m}_i$ is the prior and $M$ is an arbitrary constant mass, which we set as $M=10^{12} M_{\odot}$. We set the prior $\hat{m}_i=M_{tot,ini}/N$, where $M_{tot,ini}$ is the initial total mass of the model system, and $N$ is the number of particles in the model. As with Papers 1 and 2, we write $\epsilon=\epsilon'\epsilon''$, where
\begin{equation}
\epsilon''=\frac{10}{\text{max}_i\left(M\sum_j\frac{W(r_{ij},h_j)}{\rho_{t,j}}\Delta_{\rho_j}(t)\right)}.
\end{equation}

Following \cite{DeL08}, we use temporally smoothed versions of $\Delta_{\rho_j}$, $\hat{\mathcal{L}}$ and $\rho_j$. As opposed to the fixed values of the velocity error, $\sigma_{x,j}$, which were used in Paper 2, we now use values based on $Gaia$'s performance estimates as discussed in Section \ref{Error}. In other words, we take into account the difference in errors among different velocity components for different target stars.

We have again performed a parameter search for the optional parameters as demonstrated in Paper 1. These parameters are $\epsilon'$, which controls the balance between speed and smoothness, $\mu$, which controls the level of regularisation, $\alpha$, which controls the degree of temporal smoothing and $\zeta$, which controls the magnitude of the velocity observables contribution to the force of change. We have determined these values as $\epsilon'=0.1$, $\alpha=2.0$, $\zeta=1$ and $\mu=10^5$, these are in agreement with Paper 2. 

We calculate the angle of the bar in the model at each step. Then we rotate the model to match the bar angle of the target, assuming the bar angle is known, for the purposes of calculating the observables in the same reference frame. Paper 2 demonstrates that this method will allow the pattern speed to be recovered along with the density and velocity profiles. When applying this to the Milky Way we will not know the exact bar angle, however here, we assume that the bar angle is known for simplicity. 


\section{Results}
\label{R}
In this section we present the results from our models using \sc{primal}\rm. We will first show the results for the unconstrained model explained below, and then for a model where we apply \sc{primal }\rm to ideal data, i.e. the position and velocities are measured with no error. Then we show our fiducial model where \sc{primal }\rm is applied to the error added data ignoring dust extinction. Then we demonstrate the importance of using all three dimensions of the velocity constraints, and the importance of calculating density using stars with $V>14.5$ mag. We then show models with different initial conditions. Table \ref{Mpars} shows a summary of the models including the bar pattern speeds, the likelihood values $\mathcal{L}_{v_r}$, $\mathcal{L}_{v_{\alpha}}$ and $\mathcal{L}_{v_{\delta}}$ where
\begin{equation}
\mathcal{L}=\sum_j\ln\left(\frac{\hat{\mathcal{L}}_j}{\rho_j}\right),
\end{equation}
and the $\chi^2_{\rho}$ for the density, where
\begin{equation}
\chi^{2}_{\rho} = \frac{\sum \Delta_{\rho}^{2}}{N_r}.
\label{chi2}
\end{equation}
Note that we include only target particles with $V\leq14.5$ mag and $d_{obs}\leq10$ kpc, and $N_r$ is the number of particles satisfying this criteria. Note that although we seek to maximise likelihood, the values in Table \ref{Mpars} are $-\mathcal{L}$, and hence smaller values mean higher likelihood. Note that as discussed in Section \ref{M2M}, we do not take into account the error in density. Especially for distant target stars, the density tends to be overestimated, because of the larger errors in the distance, and therefore $\chi^2_{\rho}$ is unlikely to be a fair measurement of the goodness of fit.

\subsection{Unconstrained Model}
Firstly we show Model $i$, where all the constraints from M2M modelling have been turned off and the system is merely allowed to evolve within its own self-gravity and the fixed potential of the dark matter halo. Model $i$ is for reference and comparison with the other models with M2M modelling, as the known dark matter halo and the similar initial condition of the model to the target initial condition will contribute partially to the similar mass distribution and kinematics of the final model system to those of the target system.

Fig. \ref{AG} shows the radial profiles of the surface density, $\Sigma$, the radial, $\sigma_r$, and vertical, $\sigma_z$, velocity dispersion and the mean rotational velocity, $v_{\text{rot}}$, for the target (black solid) and Model $i$ (green dash) compared to the initial model (blue dot). The unconstrained model does not well reproduce the target in most areas. The $\Sigma$ profile shows an overestimation of the density within 9 kpc. This is unsurprising due to the lower scale length of the initial model disc. The $\sigma_r$ and $\sigma_z$ profiles match poorly within 5 kpc of the centre. However they are reproduced nicely in the outer regions, without the help of \sc{primal}\rm. The $v_{\text{rot}}$ profile is overestimated across the entire disc because of the higher surface density in the inner region. Fig. \ref{iAB} shows the fractional difference between the target and Model $i$ (green dash) in the radial profiles for comparison with the other models. The fractional surface density difference is given by
\begin{equation}
\Delta\Sigma=(\Sigma-\Sigma_t)/\Sigma_t,
\end{equation}
where $\Sigma_t$ is the true surface density of the target, and a similar equation is used for evaluating the fractional velocity errors in Fig \ref{iAB}.

The top middle panel of Fig. \ref{12} shows the fractional surface density difference between the target (top left panel) and Model $i$ in a face-on view. The fractional difference in the surface density map of the model and the target are calculated using the cloud in cell method on a 240 by 240 grid. Fig. \ref{12} shows a substantial overdensity in the model within $R_G\approx6$ kpc. This is to be expected, as without constraints from \sc{primal}\rm, the model disc remains more centrally concentrated than the target due to the initial smaller scale length of 2 kpc.

We have measured the pattern speed of the bar of the target galaxy by measuring the difference in the bar angle at different epochs. The bar pattern speed measured is $\Omega_{t,p} = 28.9 \text{ km s}^{-1}\text{kpc}^{-1}$ for the target galaxy. The pattern speed of the bar for Model $i$ is overestimated significantly with $\Omega_p = 34.3 \text{ km s}^{-1}\text{kpc}^{-1}$ for Model $i$.

\subsection{Ideal Data}
In this section we show Model A which contains no error in the target data for reference. This is similar to Model D from Paper 2, which uses the same target galaxy and initial conditions for the model. In this paper we use a more realistic selection of the target data, i.e. $V\leq14.5$ mag (corresponding to $d_{obs}\approx9.55$ kpc for M0III stars), compared with $R_G\leq10$ kpc used in Paper 2, and utilize observables in equatorial coordinates as discussed in Section \ref{Error}. A more detailed study of \sc{primal }\rm when applied to data with no error is the subject of Paper 2.

\begin{figure}
\centering
\resizebox{\hsize}{!}{\includegraphics{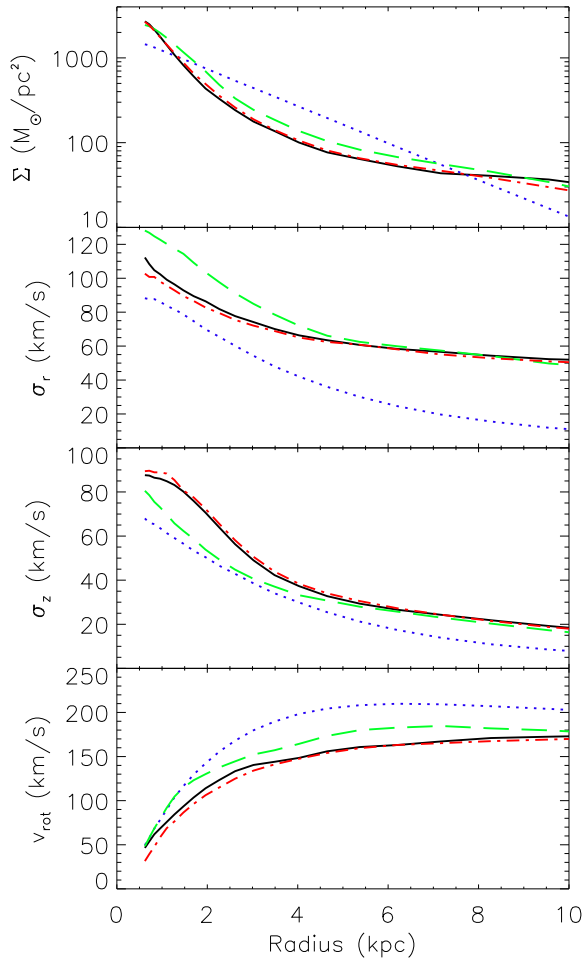}}
\caption{Surface density profile (upper), radial velocity dispersion (upper middle), vertical velocity dispersion (lower middle) and rotation velocity (lower) for the Initial model (blue dot), target (black solid), Model A (red dash-dot) and Model $i$ (green dash).}
\label{AG}
\end{figure} 

Fig. \ref{AG} shows the radial profiles of the surface density, $\Sigma$, the radial, $\sigma_r$, and vertical, $\sigma_z$, velocity dispersion and the mean rotational velocity, $v_{\text{rot}}$, for the target (black solid) and Model A (red dash-dot) compared to the initial model (blue dot). As in Papers 1 and 2, these radially binned profiles are not directly constrained by \sc{primal}\rm, but are reproduced remarkably well, especially if compared with the unconstrained model, Model $i$. Fig. \ref{AG} shows a substantial increase in the radial velocity dispersion and a corresponding decrease in the mean rotational velocity from the initial to the final model, leading to an excellent agreement with the target profiles in all areas apart from the inner 3 kpc of $\sigma_r$ and $v_{\text{rot}}$ which are slightly underestimated, corresponding to the boxy structure. Fig. \ref{iAB} shows the fractional difference between the radial profiles of the target and Models $i$ and A. Model A (red dash-dot) shows less than ten percent error in all areas apart from the outer edge of the density profile and the inner 1 kpc in the rotation velocity profile.

The top right panel of Fig. \ref{12} shows an excellent recovery of the face-on view of the surface density distribution in the middle region of Model A. The recovery is still flawed however, including a ring of underdensity around $r=10$ kpc, which is due to the failure to recover the spiral/ring structure, which is seen in the target galaxy in the top left panel (see also Fig. \ref{OT}). The pattern speed of the bar is recovered extremely well however with $\Omega_p = 28.5 \text{ km s}^{-1}\text{kpc}^{-1}$ for the final model compared to the target of $\Omega_{t,p} = 28.9 \text{ km s}^{-1}\text{kpc}^{-1}$ (see Table \ref{Mpars}). Additionally Fig. \ref{OT} shows the morphology of Model A reproduces well the boxy morphology of the Target's central bulge. The values of $\chi^2$, $\mathcal{L}_{v_r}$, $\mathcal{L}_{v_{\alpha}}$ and $\mathcal{L}_{v_{\delta}}$ from Model A (shown in Table \ref{Mpars}) are all better than those for Model $i$, the unconstrained model. They cannot be directly compared to the results for subsequent models because the positions of the tracers will have changed and different tracers may have been selected for use by the $d\leq10$ kpc and $V\leq14.5$ selection criteria.

\subsection{Fiducial Model}
\begin{figure}
\centering
\resizebox{\hsize}{!}{\includegraphics{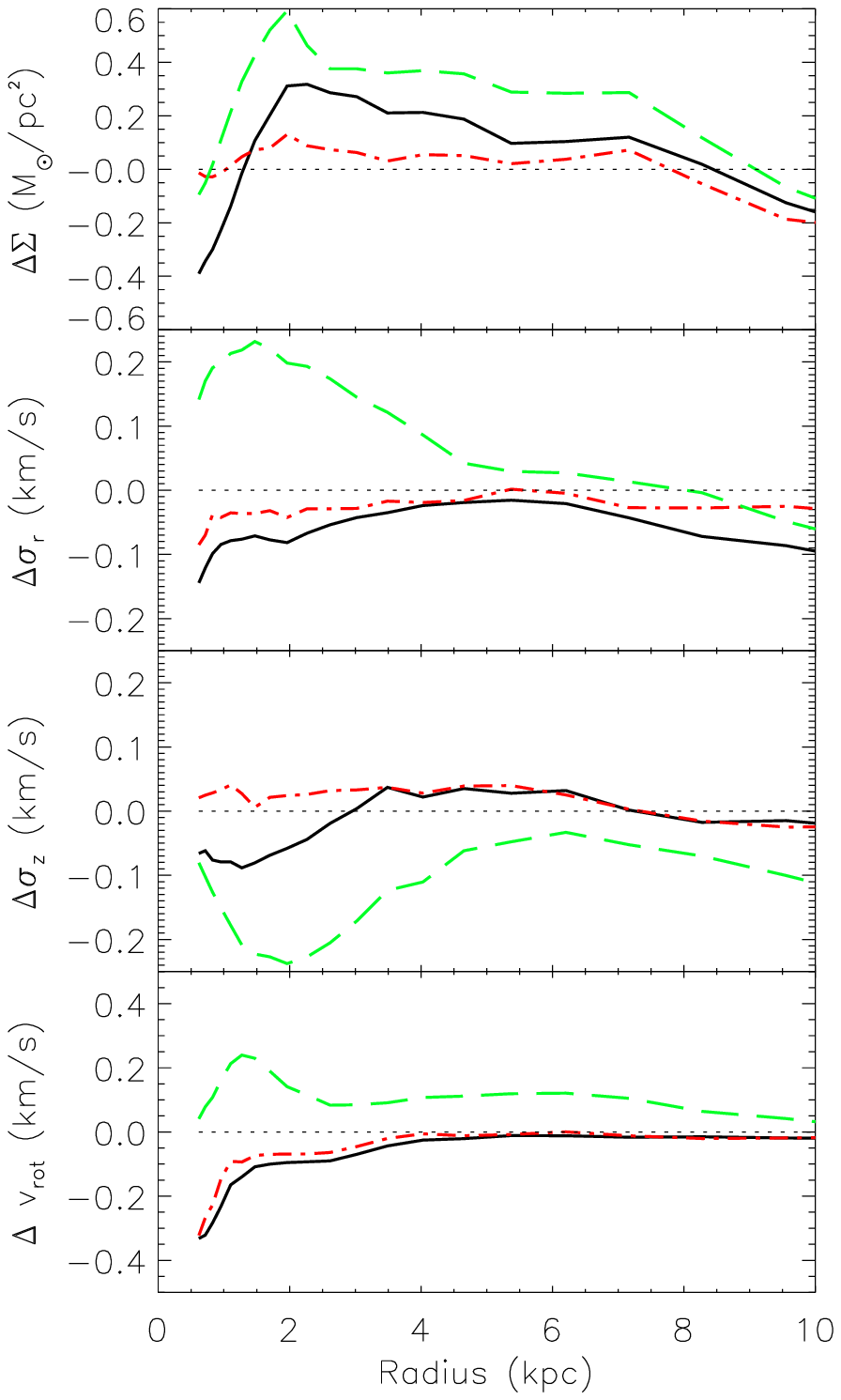}}
\caption{Fractional difference between models and target in the radial profile of the surface density (upper), radial velocity dispersion (upper middle), vertical velocity dispersion (lower middle) and rotation velocity (lower) for Model i (green dash), A (red dash-dot) and B (black solid).}
\label{iAB}
\end{figure}

In this section we present Model B, our model which best reproduces the target galaxy described in Section \ref{Error} when working with the error added observables. Fig. \ref{iAB} shows the fractional difference in the radial profiles for Model B (black solid) compared with the target galaxy. The final profiles reproduce the target profiles reasonably well, considering the parallax errors present in the observational data. There is however a noticeable decrease in accuracy when compared with Model A (red dash-dot). There is an overestimation of the density between $R_G\approx2$ and 4 kpc, and an underestimation within 1 kpc. There is also an underestimation in the inner regions of the $\sigma_r$, $\sigma_z$ and $v_{\text{rot}}$ profiles. This drop in accuracy is to be expected due to the addition of observational error. The inaccuracy in the surface density profile is believed to be due to systematic error in the density estimate of the target galaxy as we see in Fig. \ref{DenError}. The error in the density estimate is discussed further in Section \ref{DS}.

The left panel of the 2nd row of Fig. \ref{12} shows that there is increased overestimation of the density except in the bar region in Model B when compared with Model A (top right). This matches what is seen in Fig. \ref{iAB}, with the overestimation greatest between $R_G\approx2$ and 4 kpc, and an underestimation present in the central 1 kpc.

Table \ref{Mpars} shows a pattern speed of the bar of $\Omega_p = 28.6 \text{ km s}^{-1}\text{kpc}^{-1}$ for Model B, compared to $\Omega_{t,p} = 28.9 \text{ km s}^{-1}\text{kpc}^{-1}$ for the target. This is a remarkably good recovery considering the less accurate constraints in the inner region of the target galaxy and considering our naive application of \sc{primal }\rm to the error added data, and is encouraging for further development.

\subsection{Limited velocity constraints}
In this section we show the importance of using velocity constraints, as opposed to merely density constraints. We also show the importance of using three dimensional velocity constraints, as using either $v_r$ or $v_{\alpha,\delta}$ alone results in an inferior model. 

\begin{figure*}
\centering
\includegraphics[width=0.97\hsize]{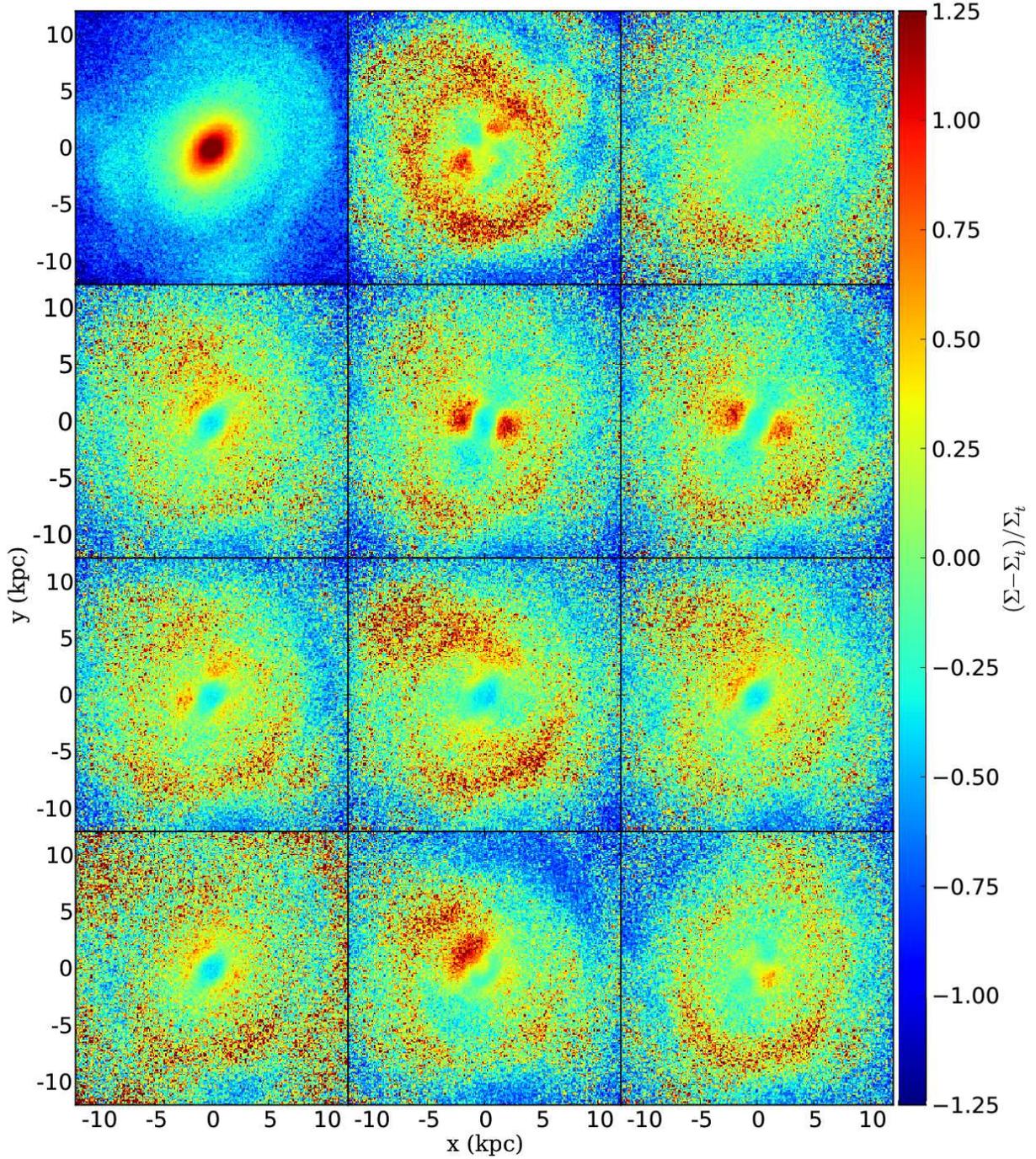}
\caption{Face-on surface density map of the target (top left) and the face-on fractional projected surface density difference maps ($\Delta\Sigma=(\Sigma-\Sigma_t)/\Sigma_t,$) calculated with a cloud-in-cell method on a 240$\times$240 grid, for Models $i$ (top middle), A (top right), B (2nd row left), C (2nd row middle), D (2nd row right), E (3rd row left), F (3rd row middle), G (3rd row right), H (bottom left), I, (bottom middle) and J (bottom right) plotted for comparison. The difference maps use the same scale as given by the colour bar. Red shows an overdensity in the model, and blue is an underdensity in the model. The surface density of the target (top left) uses its own logarithmic colour scale.}
\label{12}
\end{figure*}

\begin{figure}
\centering
\resizebox{\hsize}{!}{\includegraphics{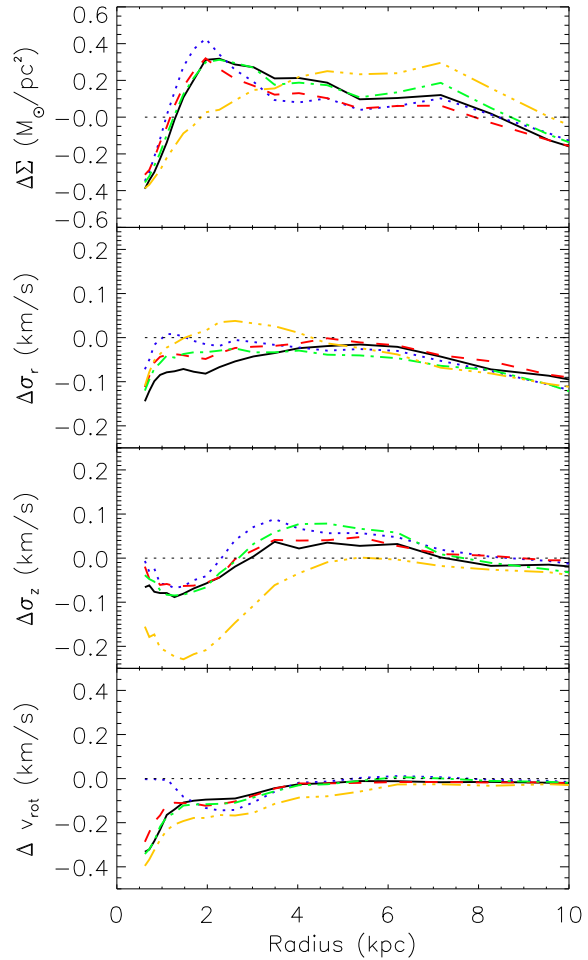}}
\caption{Same as Fig. \ref{iAB}, but for Model B (black solid), Models C (blue dot), D (red short-dash) and E (green dash-dot) which use only the density, or specific velocities as constraints and Model F (yellow triple-dot-dash) which only calculates density from stars with $V<14.5$.}
\label{BCDEF}
\end{figure}
\begin{figure}
\centering
\resizebox{\hsize}{!}{\includegraphics{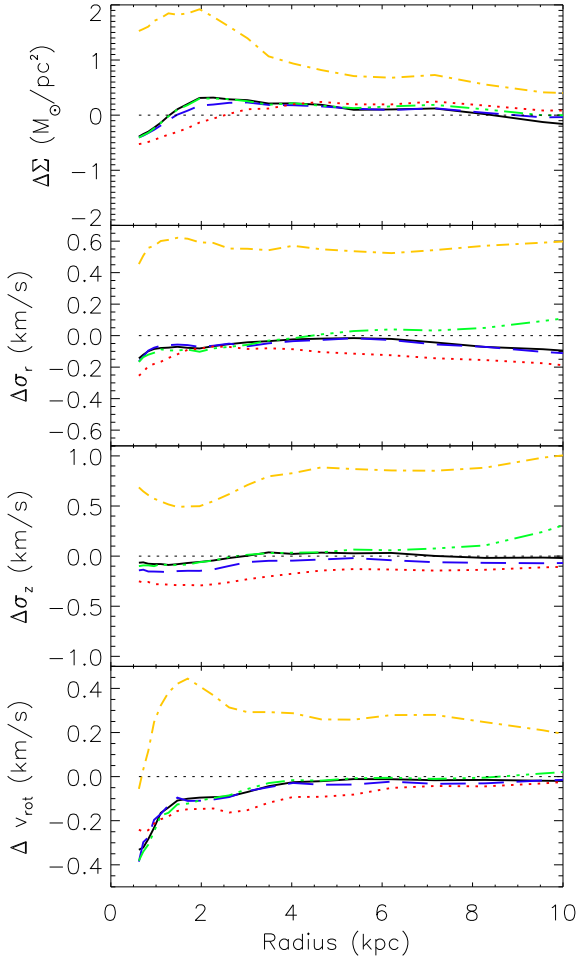}}
\caption{Same as Fig. \ref{iAB}, but for Model B (black solid), Model G (blue dash) which has $R_{d,ini}=4.0$ kpc, G$_i$ (red dot) which is Model G without constraints, Model H (green triple-dot-dash) which has $M_{d,ini}=10^{11}M_{\odot}$ and H$_i$ (yellow dash-dot) which is Model G without constraints. Note the scale for this figure is different to that of Figs. \ref{iAB}, \ref{BCDEF} and \ref{BIJ}.}
\label{BGH}
\end{figure}
\begin{figure}
\centering
\resizebox{\hsize}{!}{\includegraphics{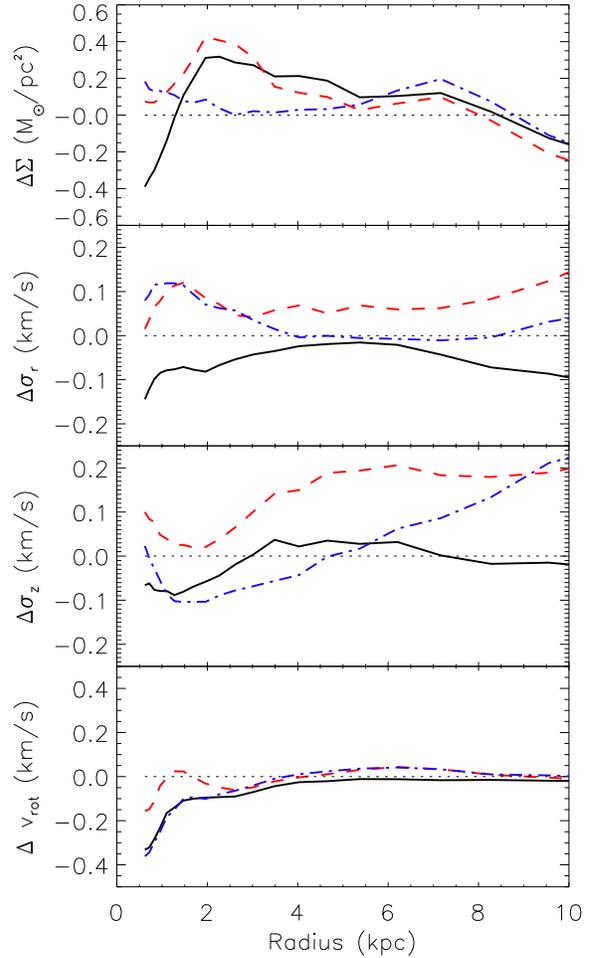}}
\caption{Same as Fig. \ref{iAB}, but for Model B (black solid), Model I (red dash) which uses M0III tracers with extinction added and Model J (blue dash-dot) which uses RC tracers with extinction added.}
\label{BIJ}
\end{figure}

Fig. \ref{BCDEF} shows the fractional difference in the radial profiles for Model C (blue dot), performed using only the density observables as constraints. Because the density is directly linked to the positions of the target stars, the error in the density observables can become quite high as you get further from the Sun, as shown in the top right panel of Fig. \ref{DenError}. The top left panel of Fig. \ref{DenError} shows that the density in the inner region of the target galaxy is overestimated. As a result, the recovery of the density around 2 kpc is worse than the fiducial model, Model B (black solid). Fig. \ref{BCDEF} also shows the $\sigma_z$ profile is a better match to the target in the inner 2 kpc but worse around 4 kpc when compared to the fiducial model. The $v_{\text{rot}}$ profile is better reproduced at 0.5 kpc, but is worse around 2.5 kpc. This is unsurprising as there are no constraints upon the velocity. Interestingly, we find an improvement in the $\sigma_r$ profile in the central part of the galaxy. However, we believe this is a coincidence and higher $\sigma_r$ is driven by overestimated density constraints.

The area of overestimated density can be clearly seen in Fig. \ref{12}, which shows the fractional surface density difference map. The middle panel in the 2nd row of Fig. \ref{12} shows two patches of substantial overestimation either side of the bar in Model C. This is because Model C contains a substantial bulge but a weak bar. The pattern speed of the bar recovered is worse than in Model B with $\Omega_p = 26.1 \text{ km s}^{-1}\text{kpc}^{-1}$ for Model C, compared to the target of $\Omega_{t,p} = 28.9 \text{ km s}^{-1}\text{kpc}^{-1}$.

Fig. \ref{BCDEF} shows the fractional differece in the radial profiles for Model D (red dash), performed using density and radial velocity observables as constraints. When comparing Model D with Model C (blue dot), we see an improvement in the $\Sigma$ profile and the velocity profiles, apart from the inner 2 kpc of the $\sigma_r$ and $v_{\text{rot}}$ profiles. Fig. \ref{12} shows the fractional surface density difference map for Model D (2nd row right), which when compared with Model C (2nd row middle) shows very little difference. The pattern speed of the bar for Model D however, has become worse when compared with Model C, with $\Omega_p = 25.0 \text{ km s}^{-1}\text{kpc}^{-1}$ for Model D, compared to the target of $\Omega_{t,p} = 28.9 \text{ km s}^{-1}\text{kpc}^{-1}$. Therefore we think that it is important to include the proper motions as constraints.

Fig. \ref{BCDEF} also shows the fractional difference in the radial profiles for Model E (green dash-dot), performed using density and proper motion observables as constraints. When comparing Model E and Model B (black solid), we see that using the proper motion constraints only rather than the full velocity constraints has improved the recovery of the $\sigma_{r}$ profile within 3 kpc. It has however resulted in a worse recovery of the $\sigma_z$ profile. The $v_{\text{rot}}$ and $\Sigma$ profiles in general remain similar.

The left panel of the 3rd column of Fig. \ref{12} shows the fractional surface density difference map for Model E, which when compared with Model C (2nd row middle) shows a noticeable improvement: the areas of overdensity on either side of the bar have been removed, and the recovery is more similar to Model B (2nd row left). The pattern speed of the bar for Model E is however worse when compared with Models B or C (see Table \ref{Mpars}), with $\Omega_p = 33.8 \text{ km s}^{-1}\text{kpc}^{-1}$ for Model E, compared to the target of $\Omega_{t,p} = 28.9 \text{ km s}^{-1}\text{kpc}^{-1}$.

When we compare Models C, D and E with Model B, we find Model B to be superior, especially when aspiring for an accurate recovery of the pattern speed of the bar, leading us to conclude that the three dimensional velocity information is an important constraint to use when it is available. This agrees with our findings in Paper 1, where this test was performed on data without errors. Table \ref{Mpars} shows the $\chi^2$, $\mathcal{L}_{v_r}$, $\mathcal{L}_{v_{\alpha}}$ and $\mathcal{L}_{v_{\delta}}$ for Models B, C, D and E. We see very little difference in $\mathcal{L}_{v_{\alpha}}$ and $\mathcal{L}_{v_{\delta}}$, however the values of $\mathcal{L}_{v_r}$ show the best recovery of the radial velocities is actually found by Model C, the model which only uses the density constraint. We find this odd, but it matches what we observe in Fig. \ref{BCDEF}, and as we discussed above, this is just a coincidence due to the overestimation of the density.

\subsection{The importance of the data selection}
\label{DS}
As discussed in Section \ref{Error}, we use only part of the data available to us as constraints to avoid using the observables with too large error. In this paper we use target M0III stars with $V\leq14.5$ mag and $d_{obs}\leq10$ kpc. For the Models up to this point we do however use M0III tracers fainter than $V=14.5$ mag and $d_{obs}>10$ kpc when calculating the density of the target stars if they lie within the smoothing length $h$. Model F is constructed using the target density measured only using M0III stars with $V\leq14.5$ mag and $d_{obs}\leq10$ kpc.

Fig. \ref{BCDEF} shows the fractional difference in the radial profiles for Model F (yellow triple-dot-dash). When comparing Model F with the other models, we see the overestimation of the density at $R_G=2$ kpc present in the other models in Fig. \ref{BCDEF}, has been removed. However the $\Sigma$ profile is worse in all other areas. The three velocity profiles have all deteriorated, with the $\sigma_z$ profile being particularly poor in the inner 4 kpc and it is barely an improvement over the unconstrained case (Fig. \ref{iAB}, green dash). Fig. \ref{12} shows the fractional surface density difference in the face-on view for Model F (3rd row middle), which when compared with Model B (2nd row left), shows a substantially worse recovery with two large patches of overdensity either side of the bar. The pattern speed of the bar for Model F is the worst recovery of any of the models presented, with $\Omega_p = 22.5 \text{ km s}^{-1}\text{kpc}^{-1}$ for Model E, compared to the target of $\Omega_{t,p} = 28.9 \text{ km s}^{-1}\text{kpc}^{-1}$. This demonstrates the importance of the density measurements including faint stars. Because the density observables are different in Model F, the value of $\chi^2$ may not be directly compared to the preceding models, however the likelihoods may. The velocity likelihoods are again all very similar.

\subsection{Different initial conditions}
\label{diffini}
In this section we show Models G and H, which are performed in the same fashion as the fiducial model, Model B, but starting from different initial conditions for the model. We also show Models G$_i$ and H$_i$ which are Models G and H performed with the constraints from M2M modelling turned off.

Model G uses an intial disc with scale length $R_{d,ini}=4$ kpc, compared with the previous models which use $R_{d,ini}=2$ kpc. Fig. \ref{BGH} shows the fractional difference in the radial profiles for Model G (blue dash). When comparing Model G with the fiducial model, Model B (black solid), we see very similar $\sigma_r$ and $v_{\text{rot}}$ profiles. However the $\sigma_z$ profile is underestimated. The $\Sigma$ profile is in general superior to the fiducial model, but not by a large amount.

The right panel of the 3rd row of Fig. \ref{12} shows the fractional surface density difference map for Model G, which, when compared with Model B (2nd row left), shows almost no difference. The pattern speed of the bar for Model G is lower than the target, but still a reasonable recovery with $\Omega_p = 25.9 \text{ km s}^{-1}\text{kpc}^{-1}$, compared to the target of $\Omega_{t,p} = 28.9 \text{ km s}^{-1}\text{kpc}^{-1}$. The values of $\chi^2$ and $\mathcal{L}_{v_r}$ are slightly worse than Model B, however the values of $\mathcal{L}_{v_{\alpha}}$ and $\mathcal{L}_{v_{\delta}}$ are very similar to that of the fiducial model.

Model G$_i$ is Model G with the constraints from M2M modelling turned off. This is the same as with Model $i$, but starting from an initial disc with $R_{d,ini}=4.0$ kpc. Fig. \ref{BGH} shows the fractional difference in the radial profiles for Model G$_i$ (red dot), which when compared with Model G (blue dash) we see a worse match to the target for all the radial profiles, apart from the $\Sigma$ profile at $R\approx2.5$ kpc, which is due to the change between the underestimation in the inner region, and overestimation in the outer region. The pattern speed of the bar for Model G$_i$ is worse than for Model G, with $\Omega_p = 23.9 \text{ km s}^{-1}\text{kpc}^{-1}$, compared to the target of $\Omega_{t,p} = 28.9 \text{ km s}^{-1}\text{kpc}^{-1}$. The values of $\chi^2$, $\mathcal{L}_{v_r}$, $\mathcal{L}_{v_{\alpha}}$ and $\mathcal{L}_{v_{\delta}}$ are all worse than those of Model G.

Model H uses an initial disc with mass $M_{d,ini}=10^{11}M_{\odot}$, compared to the other models which start from a disc with $M_{d,ini}=5\times10^{10}M_{\odot}$. Fig. \ref{BGH} shows the fractional error in the radial profiles for Model H (green triple-dot-dash). When comparing Model H with Model B (black solid), we find that the result is very similar in all profiles within 5 kpc from the centre. However the fractional difference in the outer section of the profiles is significantly larger for all profiles, especially in the $\sigma_r$ and $\sigma_z$ profiles. The bottom left panel of Fig. \ref{12} shows the fractional surface density difference map for Model H, which when compared with Model B (2nd row left) shows a generally heavier disc, with overdensities present especially at large radii. This is unsurprising considering the heavier initial Model disc mass. The pattern speed of the bar for Model H is lower than the target, but still a reasonable recovery with $\Omega_p = 25.6 \text{ km s}^{-1}\text{kpc}^{-1}$ for Model H, compared to the target of $\Omega_{t,p} = 28.9 \text{ km s}^{-1}\text{kpc}^{-1}$. Similarly to Model G, The values of $\chi^2$ and $\mathcal{L}_{v_r}$ are slightly worse than Model B, however the values of $\mathcal{L}_{v_{\alpha}}$ and $\mathcal{L}_{v_{\delta}}$ are very similar to that of the fiducial Model.

Model H$_i$ is Model H with the constraints from M2M modelling turned off. This is the same as with model $i$ but starting from an initial disc with $M_{d,ini}=10^{11}M_{\odot}$. Fig. \ref{BGH} shows the fractional difference in the radial profiles for Model H$_i$ (yellow dash-dot), which when compared with Model H (green triple-dot-dash) shows a very poor recovery of all the radial profiles. The pattern speed of the bar for Model H$_i$ is substantially overestimated with $\Omega_p = 44.4 \text{ km s}^{-1}\text{kpc}^{-1}$, compared to the target of $\Omega_{t,p} = 28.9 \text{ km s}^{-1}\text{kpc}^{-1}$. The values of $\chi^2$, $\mathcal{L}_{v_r}$, $\mathcal{L}_{v_{\alpha}}$ and $\mathcal{L}_{v_{\delta}}$ are all significantly worse than those of Model H.

Models G and H show that the results from \sc{primal }\rm are not heavily dependent on the initial conditions of the model. As with most modelling methods, it is easier to recover the target properties if the initial model is close to the target. In the final application we can iteratively change the initial condition, and find a suitable one. Comparing Models G and H with Models G$_i$ and H$_i$ show that \sc{primal }\rm is able to recover the properties of the target galaxy from initial discs which would otherwise not evolve into a galaxy which resembles the target. In particular the comparison between Model H and Model H$_i$ shows the power of \sc{primal }\rm to recover the properties of the target galaxy from a model which is initially very different from the desired solution.

\section{Dust extinction}
\label{Ex}
In previous sections, we applied \sc{primal }\rm to the mock data constructed without the dust extinction for simplicity, and to highlight the effect of the observational errors on the modelling of the Galactic disc. However, in the real galaxy, there is the dust extinction which changes the brightness and the colours of the stars, and can block their light completely depending on their distance and the position in the sky.

Interstellar extinction is a major problem that must be addressed before a convincing model of the Milky Way can be produced. Unlike surveys of external galaxies, where the Galactic extinction can be corrected for with a function $A_\lambda(l,b)$ \citep[e.g.][]{SFD98}, we need three dimensional extinction models, e.g. a function $A_\lambda(l,b,d)$, where $d$ is the distance from the Sun. While there are three dimensional extinction maps, they do not cover the entire sky, for example the map by \cite{DS01}, fitted to the far-infrared (FIR) and NIR data from the COBE/DIRBE instrument for Galactic latitudes $\mid b \mid \leq 30^{\text{o}}$ and $\mid l\mid \leq 20^{\text{o}}$. Another example is shown in \cite{MRRSP06}, for $\mid l \mid \leq 100^{\text{o}}$ and $\mid b \mid \leq 10^{\text{o}}$. However a continuous estimate of $A_\lambda(l,b,d)$ has not yet been constructed \citep{RB13}. Ways to constrain the extinction on any one star can be determined however, using a Bayesian method \citep[e.g.][]{BJ11}, and a method using the Two Micron All Sky Survey (2MASS) near infra-red (NIR) and Spitzer-IRAC mid infra-red (MIR) photometry called the Rayleigh-Jeans Colour Excess (RJCE) method \citep{MZN11}. The RJCE method works by comparing changes in stellar NIR-MIR colours due to interstellar reddening which can be calculated as stars are all essentially the same colour in the Rayleigh-Jeans part of the spectral energy distribution. \cite{NZM12} have used the RJCE method to produce a 2D map of extinction in the Galactic mid-plane for $256^{\text{o}} < l < 65^{\text{o}}$ and $\mid b\mid\leq1^{\text{o}}-1.5^{\text{o}}$ (with $\mid b\mid\leq4^{\text{o}}$ for certain longitudes), up to $d\approx8$ kpc.

To add extinction to our target tracers, we use the extinction map of the Milky Way taken from Galaxia \citep{SBJB11}. The publicly available population synthesis code, Galaxia, generates stellar populations from a galaxy model. Galaxia uses a 3D polar logarithmic grid of the dust extinction which is constructed from the method described by \cite{BKF10} and using the dust maps from \cite{SFD98}. We calculate extinction values for our target for each individual M0III tracer. We then modify the magnitudes and colours of the tracers based upon the extinction and apply the $Gaia$ expected error as shown in Section \ref{Error}.

In this section we demonstrate how \sc{primal }\rm performs when applied to the mock data considering the dust extinction. We first show Model I, which uses the M0III tracers used in the preceding models, with dust extinction added to our mock data. Then we show Model J, which uses RC stars with assumed $M_V=1.27$ and $V-I_c=1.0$ as tracers, with dust extinction added in the same fashion.

Fig. \ref{DE} shows real versus observed distance for M0III stars with extinction (middle) and RC stars with extinction (lower). The middle panel of Fig \ref{DE} shows that the accuracy within 4 kpc remains excellent even with the addition of extinction to our M0III tracers, however there is a large drop in accuracy further than this. It is encouraging however that the highest concentration of particles remains centred around the 1:1 line. The bottom panel of Fig. \ref{DE} shows a large spread of accuracies for the fainter RC tracers. In this first investigation we set a selection limit of $V\leq16.5$ and $d_{obs}\leq10$ kpc for the models with extinction to increase the number of sampled stars, deferring an extensive investigation into the selection criteria to following work.

Fig. \ref{Exmap} shows the face-on (upper panels) and edge-on (lower panels) distribution of M0III stars with error but without extinction (2nd column), M0III stars with extinction (3rd column) and RC stars with extinction (right). A comparison of the 2nd and 3rd column panels of Fig. \ref{Exmap} shows that the addition of extinction has a substantial effect on the amount of data available at the galactic centre, with the data in the plane being lost from $d_{obs}\approx3$ kpc towards the galactic centre. The right panels of Fig. \ref{Exmap} show that for the RC tracers, the $V\leq16.5$ mag limit leaves only a small amount of target data available to use as constraints. There is no evidence of an overdensity from the galactic centre, and a large amount of data has been lost from the galactic plane.

Fig. \ref{DenError} shows the fractional density error of the mock data against observed Galactocentric radius (left) and observed distance from the Sun (right) for M0III stars without extinction (upper), M0III stars with extinction (middle) and RC stars with extinction (lower). The middle left panel of Fig. \ref{DenError} shows a similar trend to the case without extinction (upper left), however the worst overestimation of the density is now spread between $R_G\approx2$ and 4 kpc. The lower panels of Fig. \ref{DenError} show an even larger spread of the overestimation between $R_G\approx3$ and 7 kpc, and the density for stars whose observed distance is more than 6 kpc is mostly underestimated.

Fig. \ref{BIJ} shows the fractional difference in the radial profiles for Model I (red dash) which uses M0III tracers with dust extinction and observational error. Model I shows a substantial overestimation of the density around 2 kpc, a general overestimation of the $\sigma_r$ and $\sigma_z$ profiles, but a better recovery in the inner region of the $v_{\text{rot}}$ profile. The bottom middle panel of Fig. \ref{12} shows the fractional surface density difference map for Model I, which when compared with Model B (2nd row left) shows a substantially worse recovery. There is an overdensity near the centre, which is not present in the fiducial model, and a large underdensity in the top right of the plot. The pattern speed of the bar is again recovered well however with $\Omega_p = 27.7 \text{ km s}^{-1}\text{kpc}^{-1}$ for Model I, compared to the target of $\Omega_{t,p} = 28.9 \text{ km s}^{-1}\text{kpc}^{-1}$. The density overestimation in the inner part of the $\Sigma$ profile and in Fig. \ref{12} is concerning, although it is not surprising considering the overestimation shown at $R_G\approx2$-3 kpc in the middle left panel of Fig. \ref{DenError}. Due to the extinction, the number of target stars selected has decreased dramatically from 517,527 to 173,821 (see Table \ref{Mpars}) and the location of the remaining observables will have moved. Therefore in Model I the values of $\chi^2$, $\mathcal{L}_{v_r}$, $\mathcal{L}_{v_{\alpha}}$ and $\mathcal{L}_{v_{\delta}}$ may not be directly compared to the preceding models.

Fig. \ref{BIJ} shows the fractional difference in the radial profiles for Model J (blue dash-dot) which uses RC tracers with dust extinction and error added to the target data. Model J shows a good recovery of the $\Sigma$ profile. Model J is also better than model B (black solid) between $R_G=2$ and 6 kpc, which is very encouraging. Model J is similar to Model I in the inner 2 kpc of the $\sigma_r$ profile, and is a substantially better reproduction of the rest of the profile, again superior to Model B. The $\sigma_z$ profile for model J is odd, with a substantial underestimation in the inner region, and a substantial overestimation in the outer region. The $v_{\text{rot}}$ profile is similar to that of Model B.

The bottom right panel of Fig. \ref{12} shows the fractional surface density difference map for Model J, which when compared with Model I (bottom middle) shows a better recovery, although it is still noticeably worse than Model B (2nd row left). The pattern speed of the bar is again recovered well with $\Omega_p = 27.3 \text{ km s}^{-1}\text{kpc}^{-1}$ for Model J, compared to the target of $\Omega_{t,p} = 28.9 \text{ km s}^{-1}\text{kpc}^{-1}$. We find the accuracy of Model J to be very encouraging for our future exploration of more realistic mock data containing multiple populations. Due to the use of RC tracers the number of selected tracers in Model J (52,111) has again decreased, and thus the values of $\chi^2$, $\mathcal{L}_{v_r}$, $\mathcal{L}_{v_{\alpha}}$ and $\mathcal{L}_{v_{\delta}}$ may not be directly compared to the preceding models.

The level of accuracy of Models I and J is still encouraging, considering the amount of information which is lost due to extinction. We find it surprising however that the RC tracers lead to a more accurate model. However, self-gravity leads to a stable model in a non-linear way, and different constraints sometimes act counter-intuitively. We stress the need for further testing with mock data, and the selection criteria used. This is what we will explore thoroughly in the next stage of our work. What we can conclude for the time being from this initial trial is that the accuracy of the recovery is difficult to control for M2M modelling, and a careful balance must be reached between the quantity and quality of data which are used for observables. The data selection criteria will need to be different depending on the type of star. We do not consider it useful to do extensive testing on the selection criteria at this stage, as \sc{primal }\rm must be modified to use more realistic mock data before such tests become meaningful.

\section{Summary}
\label{SF}
We have demonstrated that \sc{primal }\rm can recover to a reasonable degree the properties of a target disc system with a bar/boxy structure in a known dark matter halo potential despite the presence of error in the observational data. To allow us to do this we have modified \sc{primal }\rm to use equatorial coordinates which is the form of data $Gaia$ will provide. In this paper the error added  observables are compared with the model at the observed position of the target particles, and the masses of the model particles are altered to reproduce the target observables. The gravitational potential is calculated self-consistently to allow the potential to change along with the model. We have demonstrated that \sc{primal }\rm can recover the pattern speed of the bar to an excellent degree under these conditions.

We stress that this paper is a first attempt at dynamical modelling taking into account the $Gaia$ error, and is used as a demonstration of how we can and will deal with this, not a statement of the final capability or accuracy of the algorithm. It is however encouraging that the $Gaia$ errors are good enough to recover galactic structure, at least with this simple model, and is worth further exploration of this methodology. We are aware however that this is still a simplified case containing many assumptions. In a forthcoming work we will further explore the effect of extinction modifying \sc{primal }\rm to work with more realistic mock observational data which will consist of multiple stellar populations. A strong assumption made at this stage is that we assume the relationship between cluster mass and the number density of M0III stars is known. This is of course not the case, and will have to be addressed in further works. Additionally this paper assumes a known dark matter halo potential for simplicity, whereas in reality the dark matter distribution of the halo remains very much unknown. The halo does however have a significant effect on the dynamics of the galaxy, and thus we intend to explore different dark matter halo density profiles in future work including the possibility of using a live halo.

We remain optimistic for the ongoing development of \sc{primal}\rm, and continue to develop a unique tool to recover the dynamical properties of the Milky Way from the future large-scale stellar survey data.

\section*{Acknowledgements}
We thank an anonymous referee for their careful review of the manuscript. The calculations for this paper were performed on Cray XT4 at Center for Computational Astrophysics, CfCA, of the National Astronomical Observatory of Japan and the DiRAC facilities (through the COSMOS consortium) jointly funded through STFC and BIS. The authors acknowledge the use of the IRIDIS High Performance Computing Facility, and associated support services at the University of Southampton. We would also like to thank PRACE for the use of the Cartesius facility. This work was carried out, in part, through the $Gaia$ Research for European Astronomy Training (GREAT-ITN) network. The research leading to these results has received funding from the European Union Seventh Framework Programme ([FP7/2007-2013] under grant agreement number 264895). We would also like to thank Francesca Figueras and Merc\`{e} Romero-G\'{o}mez for providing the subroutine to calculate the $Gaia$ performance errors, Sanjib Sharma for providing the Galaxia extinction maps and Saulius Raudeliunas for catching an important typo in the paper.

\bibliographystyle{mn2e}
\bibliography{ref2}

\begin{thebibliography}{51}
\expandafter\ifx\csname natexlab\endcsname\relax\def\natexlab#1{#1}\fi

\bibitem[{{Allende Prieto} {et~al}\mbox{.}(2013){Allende Prieto}, {Koesterke},
  {Ludwig}, {Freytag}, \& {Caffau}}]{APetal13}
{Allende Prieto} C., {Koesterke} L., {Ludwig} H.-G., {Freytag} B., {Caffau} E.,
  2013, \aap, 550, A103

\bibitem[{{Bahcall} \& {Soneira}(1980)}]{BS80}
{Bahcall} J.~N., {Soneira} R.~M., 1980, \apjs, 44, 73

\bibitem[{{Bailer-Jones}(2011)}]{BJ11}
{Bailer-Jones} C.~A.~L., 2011, \mnras, 411, 435

\bibitem[{{Binney}(2012)}]{B12-1}
{Binney} J., 2012, \mnras, 426, 1328

\bibitem[{{Bissantz} {et~al}\mbox{.}(2004){Bissantz}, {Debattista}, \&
  {Gerhard}}]{BDG04}
{Bissantz} N., {Debattista} V.~P., {Gerhard} O., 2004, \apjl, 601, L155

\bibitem[{{Bland-Hawthorn} {et~al}\mbox{.}(2010){Bland-Hawthorn}, {Krumholz},
  \& {Freeman}}]{BKF10}
{Bland-Hawthorn} J., {Krumholz} M.~R., {Freeman} K., 2010, \apj, 713, 166

\bibitem[{{Bovy} \& {Rix}(2013)}]{BR13}
{Bovy} J., {Rix} H.-W., 2013, \apj, 779, 115

\bibitem[{{Brown}(2013)}]{Brown2013}
{Brown} A.~G.~A., 2013, ArXiv e-prints

\bibitem[{{Das} {et~al}\mbox{.}(2011){Das}, {Gerhard}, {Mendez}, {Teodorescu},
  \& {de Lorenzi}}]{DGMT11}
{Das} P., {Gerhard} O., {Mendez} R.~H., {Teodorescu} A.~M., {de Lorenzi} F.,
  2011, \mnras, 415, 1244

\bibitem[{{de Bruijne}(2012)}]{dB12}
{de Bruijne} J.~H.~J., 2012, \apss, 341, 31

\bibitem[{{de Lorenzi} {et~al}\mbox{.}(2007){de Lorenzi}, {Debattista},
  {Gerhard}, \& {Sambhus}}]{DeL07}
{de Lorenzi} F., {Debattista} V.~P., {Gerhard} O., {Sambhus} N., 2007, \mnras,
  376, 71

\bibitem[{{de Lorenzi} {et~al}\mbox{.}(2008){de Lorenzi}, {Gerhard}, {Saglia},
  {Sambhus}, {Debattista}, {Pannella}, \& {M{\'e}ndez}}]{DeL08}
{de Lorenzi} F., {Gerhard} O., {Saglia} R.~P., {Sambhus} N., {Debattista}
  V.~P., {Pannella} M., {M{\'e}ndez} R.~H., 2008, \mnras, 385, 1729

\bibitem[{{Dehnen}(2009)}]{Deh09}
{Dehnen} W., 2009, \mnras, 395, 1079

\bibitem[{{Drimmel} \& {Spergel}(2001)}]{DS01}
{Drimmel} R., {Spergel} D.~N., 2001, \apj, 556, 181

\bibitem[{{Gardner} {et~al}\mbox{.}(2013){Gardner}, {Debattista}, {Robin},
  {V{\'a}squez}, \& {Zoccali}}]{GDRVZ13}
{Gardner} E., {Debattista} V.~P., {Robin} A.~C., {V{\'a}squez} S., {Zoccali}
  M., 2013, ArXiv e-prints

\bibitem[{{Grand} {et~al}\mbox{.}(2012){Grand}, {Kawata}, \& {Cropper}}]{GKC12}
{Grand} R.~J.~J., {Kawata} D., {Cropper} M., 2012, \mnras, 421, 1529

\bibitem[{{Hunt} \& {Kawata}(2013)}]{HK12}
{Hunt} J.~A.~S., {Kawata} D., 2013, \mnras, 430, 1928 (Paper 1)

\bibitem[{{Hunt} {et~al}\mbox{.}(2013){Hunt}, {Kawata}, \& {Martel}}]{HKM13}
{Hunt} J.~A.~S., {Kawata} D., {Martel} H., 2013, \mnras, 432, 3062 (Paper 2)

\bibitem[{{Jordi} {et~al}\mbox{.}(2010){Jordi}, {Gebran}, {Carrasco}, {de
  Bruijne}, {Voss}, {Fabricius}, {Knude}, {Vallenari}, {Kohley}, \&
  {Mora}}]{Jetal10}
{Jordi} C. {et~al.}, 2010, \aap, 523, A48

\bibitem[{{Katz} {et~al}\mbox{.}(2011){Katz}, {Cropper}, {Meynadier},
  {Jean-Antoine}, {Allende Prieto}, {Baker}, {Benson}, {Berthier}, {Bigot},
  {Blomme}, {Boudreault}, {Chemin}, {Crifo}, {Damerdji}, {David}, {David},
  {Delle Luche}, {Dolding}, {Fr{\'e}mat}, {Gerbier}, {Gerssen}, {G{\'o}mez},
  {Gosset}, {Guerrier}, {Guy}, {Hall}, {Hestroffer}, {Huckle}, {Jasniewicz},
  {Ludwig}, {Martayan}, {Morel}, {Nguyen}, {Ocvirk}, {Parr}, {Royer},
  {Sartoretti}, {Seabroke}, {Simon}, {Smith}, {Soubiran}, {Steinmetz},
  {Th{\'e}venin}, {Turon}, {Udry}, {Veltz}, \& {Viala}}]{Ketal11}
{Katz} D. {et~al.}, 2011, in EAS Publications Series, Vol.~45, EAS Publications
  Series, pp. 189--194

\bibitem[{{Katz} {et~al}\mbox{.}(2004){Katz}, {Munari}, {Cropper}, {Zwitter},
  {Th{\'e}venin}, {David}, {Viala}, {Crifo}, {Gomboc}, {Royer}, {Arenou},
  {Marrese}, {Sordo}, {Wilkinson}, {Vallenari}, {Turon}, {Helmi}, {Bono},
  {Perryman}, {G{\'o}mez}, {Tomasella}, {Boschi}, {Morin}, {Haywood},
  {Soubiran}, {Castelli}, {Bijaoui}, {Bertelli}, {Prsa}, {Mignot}, {Sellier},
  {Baylac}, {Lebreton}, {Jauregi}, {Siviero}, {Bingham}, {Chemla}, {Coker},
  {Dibbens}, {Hancock}, {Holland}, {Horville}, {Huet}, {Laporte}, {Melse},
  {Say{\`e}de}, {Stevenson}, {Vola}, {Walton}, \& {Winter}}]{Ketal04}
{Katz} D. {et~al.}, 2004, \mnras, 354, 1223

\bibitem[{{Kawata} \& {Gibson}(2003)}]{KG03}
{Kawata} D., {Gibson} B.~K., 2003, \mnras, 340, 908

\bibitem[{{Kawata} {et~al}\mbox{.}(2013){Kawata}, {Okamoto}, {Gibson},
  {Barnes}, \& {Cen}}]{KOGBC13}
{Kawata} D., {Okamoto} T., {Gibson} B.~K., {Barnes} D.~J., {Cen} R., 2013,
  \mnras, 428, 1968

\bibitem[{{Klypin} {et~al}\mbox{.}(2002){Klypin}, {Zhao}, \&
  {Somerville}}]{KZS02}
{Klypin} A., {Zhao} H., {Somerville} R.~S., 2002, \apj, 573, 597

\bibitem[{{Lindegren} {et~al}\mbox{.}(2012){Lindegren}, {Lammers}, {Hobbs},
  {O'Mullane}, {Bastian}, \& {Hern{\'a}ndez}}]{LLHOBH12}
{Lindegren} L., {Lammers} U., {Hobbs} D., {O'Mullane} W., {Bastian} U.,
  {Hern{\'a}ndez} J., 2012, \aap, 538, A78

\bibitem[{{Liu} {et~al}\mbox{.}(2012){Liu}, {Bailer-Jones}, {Sordo},
  {Vallenari}, {Borrachero}, {Luri}, \& {Sartoretti}}]{LBSVBLS12}
{Liu} C., {Bailer-Jones} C.~A.~L., {Sordo} R., {Vallenari} A., {Borrachero} R.,
  {Luri} X., {Sartoretti} P., 2012, \mnras, 426, 2463

\bibitem[{{Long} \& {Mao}(2010)}]{LM10}
{Long} R.~J., {Mao} S., 2010, \mnras, 405, 301

\bibitem[{{Long} \& {Mao}(2012)}]{LM12}
{Long} R.~J., {Mao} S., 2012, \mnras, 421, 2580

\bibitem[{{Long} {et~al}\mbox{.}(2013){Long}, {Mao}, {Shen}, \& {Wang}}]{LMIII}
{Long} R.~J., {Mao} S., {Shen} J., {Wang} Y., 2013, \mnras, 428, 3478

\bibitem[{{Luri} {et~al}\mbox{.}(2014){Luri}, {Palmer}, {Arenou}, {Masana}, {de
  Bruijne}, {Antiche}, {Babusiaux}, {Borrachero}, {Sartoretti}, {Julbe},
  {Isasi}, {Martinez}, {Robin}, {Reyl{\'e}}, {Jordi}, \& {Carrasco}}]{Xetal14}
{Luri} X. {et~al.}, 2014, ArXiv e-prints

\bibitem[{{Majewski} {et~al}\mbox{.}(2011){Majewski}, {Zasowski}, \&
  {Nidever}}]{MZN11}
{Majewski} S.~R., {Zasowski} G., {Nidever} D.~L., 2011, \apj, 739, 25

\bibitem[{{Marshall} {et~al}\mbox{.}(2006){Marshall}, {Robin}, {Reyl{\'e}},
  {Schultheis}, \& {Picaud}}]{MRRSP06}
{Marshall} D.~J., {Robin} A.~C., {Reyl{\'e}} C., {Schultheis} M., {Picaud} S.,
  2006, \aap, 453, 635

\bibitem[{{McMillan} \& {Binney}(2012)}]{McB12}
{McMillan} P.~J., {Binney} J., 2012, \mnras, 419, 2251

\bibitem[{{McMillan} \& {Binney}(2013)}]{MB13}
{McMillan} P.~J., {Binney} J.~J., 2013, \mnras, 433, 1411

\bibitem[{{Monaghan} \& {Lattanzio}(1985)}]{ML85}
{Monaghan} J.~J., {Lattanzio} J.~C., 1985, \aap, 149, 135

\bibitem[{{Morganti} \& {Gerhard}(2012)}]{MG12}
{Morganti} L., {Gerhard} O., 2012, \mnras, 2607

\bibitem[{{Morganti} {et~al}\mbox{.}(2013){Morganti}, {Gerhard}, {Coccato},
  {Martinez-Valpuesta}, \& {Arnaboldi}}]{MGCMA13}
{Morganti} L., {Gerhard} O., {Coccato} L., {Martinez-Valpuesta} I., {Arnaboldi}
  M., 2013, \mnras, 431, 3570

\bibitem[{{Navarro} {et~al}\mbox{.}(1997){Navarro}, {Frenk}, \&
  {White}}]{NFW97}
{Navarro} J.~F., {Frenk} C.~S., {White} S.~D.~M., 1997, \apj, 490, 493

\bibitem[{{Nidever} {et~al}\mbox{.}(2012){Nidever}, {Zasowski}, \&
  {Majewski}}]{NZM12}
{Nidever} D.~L., {Zasowski} G., {Majewski} S.~R., 2012, \apjs, 201, 35

\bibitem[{{Pasetto} {et~al}\mbox{.}(2003){Pasetto}, {Chiosi}, \&
  {Carraro}}]{PCC03}
{Pasetto} S., {Chiosi} C., {Carraro} G., 2003, \aap, 405, 931

\bibitem[{{Price} \& {Monaghan}(2007)}]{PM07}
{Price} D.~J., {Monaghan} J.~J., 2007, \mnras, 374, 1347

\bibitem[{{Rix} \& {Bovy}(2013)}]{RB13}
{Rix} H.-W., {Bovy} J., 2013, ArXiv e-prints

\bibitem[{{Robin} {et~al}\mbox{.}(2012){Robin}, {Luri}, {Reyl{\'e}}, {Isasi},
  {Grux}, {Blanco-Cuaresma}, {Arenou}, {Babusiaux}, {Belcheva}, {Drimmel},
  {Jordi}, {Krone-Martins}, {Masana}, {Mauduit}, {Mignard}, {Mowlavi},
  {Rocca-Volmerange}, {Sartoretti}, {Slezak}, \& {Sozzetti}}]{Rea12}
{Robin} A.~C. {et~al.}, 2012, \aap, 543, A100

\bibitem[{{Robin} {et~al}\mbox{.}(2003){Robin}, {Reyl{\'e}}, {Derri{\`e}re}, \&
  {Picaud}}]{Rea03}
{Robin} A.~C., {Reyl{\'e}} C., {Derri{\`e}re} S., {Picaud} S., 2003, \aap, 409,
  523

\bibitem[{{Schlegel} {et~al}\mbox{.}(1998){Schlegel}, {Finkbeiner}, \&
  {Davis}}]{SFD98}
{Schlegel} D.~J., {Finkbeiner} D.~P., {Davis} M., 1998, \apj, 500, 525

\bibitem[{{Seabroke} {et~al}\mbox{.}(2011){Seabroke}, {Prod'Homme},
  {Hopkinson}, {Burt}, {Robbins}, \& {Holland}}]{Setal11}
{Seabroke} G.~M., {Prod'Homme} T., {Hopkinson} G., {Burt} D., {Robbins} M.,
  {Holland} A., 2011, in EAS Publications Series, Vol.~45, EAS Publications
  Series, pp. 433--436

\bibitem[{{Sharma} {et~al}\mbox{.}(2011){Sharma}, {Bland-Hawthorn}, {Johnston},
  \& {Binney}}]{SBJB11}
{Sharma} S., {Bland-Hawthorn} J., {Johnston} K.~V., {Binney} J., 2011, \apj,
  730, 3

\bibitem[{{Syer} \& {Tremaine}(1996)}]{ST96}
{Syer} D., {Tremaine} S., 1996, \mnras, 282, 223

\bibitem[{{Wegg} \& {Gerhard}(2013)}]{WG13}
{Wegg} C., {Gerhard} O., 2013, \mnras, 435, 1874

\bibitem[{{Widrow} {et~al}\mbox{.}(2008){Widrow}, {Pym}, \& {Dubinski}}]{WPD08}
{Widrow} L.~M., {Pym} B., {Dubinski} J., 2008, \apj, 679, 1239

\bibitem[{{Wilkinson} {et~al}\mbox{.}(2005){Wilkinson}, {Vallenari}, {Turon},
  {Munari}, {Katz}, {Bono}, {Cropper}, {Helmi}, {Robichon}, {Th{\'e}venin},
  {Vidrih}, {Zwitter}, {Arenou}, {Baylac}, {Bertelli}, {Bijaoui}, {Boschi},
  {Castelli}, {Crifo}, {David}, {Gomboc}, {G{\'o}mez}, {Haywood}, {Jauregi},
  {de Laverny}, {Lebreton}, {Marrese}, {Marsh}, {Mignot}, {Morin}, {Pasetto},
  {Perryman}, {Pr{\v s}a}, {Recio-Blanco}, {Royer}, {Sellier}, {Siviero},
  {Sordo}, {Soubiran}, {Tomasella}, \& {Viala}}]{Wetal05}
{Wilkinson} M.~I. {et~al.}, 2005, \mnras, 359, 1306

\end{thebibliography}

\label{lastpage}
\end{document}